\documentclass[a4paper,fleqn,usenatbib,useAMS]{mnras}

\usepackage{graphicx}	
\usepackage{amsmath}	
\usepackage{amssymb}	
\usepackage{multicol}        
\usepackage{pdflscape}	
\usepackage{gensymb}
\usepackage{tikz}
\usetikzlibrary{arrows}

\newcommand{\exref}[1]{(\ref{#1})}

\renewcommand{\eqref}[1]{Eq.~(\ref{#1})}
\newcommand{\eqsref}[1]{Eqs.~(\ref{#1})}

\newcommand{\eqsand}[2]{Eqs.\ (\ref{#1}) and~(\ref{#2})}

\newcommand{\figref}[1]{Fig.~\ref{#1}}

\newcommand{\bea}{\begin{eqnarray}}
\newcommand{\eea}{\end{eqnarray}}
\newcommand{\beq}{\begin{equation}}
\newcommand{\eeq}{\end{equation}}
\newcommand{\lt}{\left}
\newcommand{\rt}{\right}

\newcommand{\tA}{\tau_{\rm A}}
\newcommand{\vA}{v_{\rm A}}

\newcommand{\tnl}{\tau_{\rm nl}}

\newcommand{\vskipfig}{\vskip-0.25cm}
\newcommand{\dd}{\partial}
\newcommand{\vB}{\mathbf{B}}
\newcommand{\vBloc}{\vB_{\rm loc}}
\newcommand{\vu}{\mathbf{u}}
\newcommand{\vb}{\mathbf{b}}
\newcommand{\vbloc}{\hat{\vb}_{\rm loc}}
\newcommand{\vz}{\mathbf{z}}
\newcommand{\vr}{\mathbf{r}}
\newcommand{\vrp}{\vr_\perp}

\newcommand{\vup}{\vu_\perp}
\newcommand{\vbp}{\vb_\perp}
\newcommand{\vzp}{\vz_\perp}

\newcommand{\dzp}{\delta z_\perp}

\newcommand{\lpar}{l_\parallel}
\newcommand{\Lpar}{L_\parallel}

\newcommand{\hlpar}{\hat{l}_\parallel}
\newcommand{\hlam}{\hat{\lambda}}
\newtheorem{conj}{Conjecture}

\usepackage[T1]{fontenc}
\usepackage{ae,aecompl}

\usepackage{txfonts}

\title[Anisotropy and intermittency in Alfv\'enic turbulence]{A statistical model of three-dimensional anisotropy and intermittency in strong Alfv\'enic turbulence}

\author[A. Mallet and A.A. Schekochihin]{A. Mallet$^{1,2}$\thanks{Contact e-mail: \href{mailto:alfred.mallet@unh.edu}{alfred.mallet@unh.edu}} and A. A. Schekochihin$^{2,3}$
\\
$^{1}$Space Science Center, University of New Hampshire, Durham, NH 03824, USA \\
$^{2}$Rudolf Peierls Centre for Theoretical Physics, University of Oxford, Oxford OX1 3NP, United Kingdom\\
$^{3}$Merton College, Oxford OX1 4JD, United Kingdom\\
}

\pubyear{2016}

\begin{document}
\label{firstpage}
\pagerange{\pageref{firstpage}--\pageref{lastpage}}
\maketitle
\begin{abstract}
We propose a simple statistical model of three-dimensionally anisotropic, intermittent, strong Alfv\'enic turbulence, incorporating both critical balance and dynamic alignment. Our model is based on log-Poisson statistics for Elsasser-field increments {\em along} the magnetic field. We predict the scalings of Elsasser-field conditional two-point structure functions with point separations in all three directions in a coordinate system locally aligned with the direction of the magnetic field and of the fluctuating fields and obtain good agreement with numerical simulations. We also derive a scaling of the parallel coherence scale of the fluctuations, $\lpar \propto \lambda^{1/2}$, where $\lambda$ is the perpendicular scale. This is indeed observed for the bulk of the fluctuations in numerical simulations.
\end{abstract}

\section{Introduction}
Turbulent plasma fills most of the visible Universe, and can be measured directly by spacecraft in the solar wind \citep{bruno2013}. In many situations, a strong mean magnetic field $\vB_0$ is present, which ensures that on scales longer than the ion gyroradius, Alfv\'enically polarized fluctuations decouple from the compressive fluctuations and satisfy the equations of reduced magnetohydrodynamics (RMHD) \citep{schektome2009}. These can be written in terms of Elsasser variables $\vzp^\pm = \vup \pm \vbp$, where $\vup$ and $\vbp$ are the velocity and magnetic-field
(in velocity units) perturbations perpendicular to the background magnetic field $\vB_0$:
\beq
\dd_t \vzp^\pm \mp v_{\rm A} \dd_z \vzp^\pm + \vzp^\mp \cdot \nabla_\perp \vzp^\pm = -\nabla_\perp p,
\label{eq:RMHD}
\eeq
where the pressure $p$ can be determined from $\nabla_\perp\cdot \vz_\perp^\pm = 0$, the Alfv\'en speed is $v_{\rm A} = |\vB_0|$, and $\vB_0$ is in the $z$ direction.

The turbulent state described by Eqs. (\ref{eq:RMHD}) is anisotropic with respect to the direction of the local magnetic field, in full MHD simulations with a strong guide field \citep[e.g.][]{oughton94,matthaeus96,matthaeus98,chovishniac,marongoldreich,bigot08}, direct numerical simulations of RMHD \citep[e.g.][]{shebalin83,oughton04,chenmallet,beresnyakanis,mallet3d} and also in the solar wind \citep[e.g.][]{horanis,podestaaniso,wicks10,chenmallet}. 
This anisotropic state can be understood on the basis of the critical-balance conjecture \citep{gs95,gs97}: the nonlinear $\tnl^\pm$ and linear $\tA^\pm\doteq \lpar^\pm/v_{\rm A}$ times should be similar at every scale in the inertial range, where $\lpar^\pm$ is the coherence length of the fluctuations along the magnetic field lines. This allows one to equate the cascade time to either of these times, and therefore, by an argument following \cite{k41}, the constancy of the energy flux through parallel scales
\beq
\epsilon^\pm \sim \frac{(\dzp^\pm)^2}{\tau_\mathrm{c}} 
\sim \frac{(\dzp^\pm)^2 v_{\rm A}}{\lpar^\pm}
\sim \text{const},
\label{eq:eps}
\eeq
implying that $(\dzp^\pm)^2 \sim \lpar^\pm (\epsilon^\pm / v_{\rm A})$, and hence a ``parallel spectral index" of $-2$, regardless of the details of the nonlinear interactions. This is, indeed, observed in the simulations and in the measurements of the solar wind cited above.

As is evident in the form of the nonlinear term in Eqs.~(\ref{eq:RMHD}), only $\vzp^\pm$ with a gradient in the direction of $\vzp^\mp$ gives rise to a nonzero contribution to the RMHD nonlinearity. Combined with the 2D-solenoidal nature of the Elsasser fields, $\nabla_\perp \cdot \vzp^\pm = 0$, this means that \emph{dynamic alignment} \citep{boldyrev,bl06} of their fluctuation vectors to within a small angle $\theta$ of each other will decrease the nonlinearity by a factor $\sin\theta$. The definition of the nonlinear time must take this into account:
\beq
\tnl^\pm \doteq \frac{\lambda}{\delta z^\mp_\perp \sin\theta},\label{eq:tnl}
\eeq
where $\lambda$ is the perpendicular coherence length. If $\theta$ depends on $\delta z^\pm_\perp$ and $\lambda$ in a non-trivial manner,  this will affect the scaling behaviour of the nonlinear time and, therefore, the scaling of the fluctuation amplitudes conditional on perpendicular scale $\lambda$, $\delta z^\pm_\perp$. Here and everywhere we will assume that the turbulence is ``balanced", i.e. $\epsilon^+ \sim \epsilon^-$ and so $\delta z_\perp^+ \sim \delta z_\perp^-$, 
$\lpar^+\sim\lpar^-$, etc.

The alignment of the fields can be linked to anisotropy of sheetlike turbulent structures within the plane perpendicular to the mean magnetic field \citep{boldyrev}. The distance field lines wander in this perpendicular plane as a result of a $\dzp^\pm$ fluctuation is
\beq
\xi \sim \lpar \frac{\dzp^\pm}{\vA}.
\label{eq:xi}
\eeq
Since $\lpar$, by definition, is the coherence length along the field line, the fluctuations must be coherent in their own direction (the ``fluctuation direction") up to a distance of at least $\xi$. However, since the fluctuation is comprised of a mixture of both Elsasser fields $\delta z^+_\perp$ and $\delta z^-_\perp$, the fluctuation direction is only defined up to the angle $\theta$ between them, and we can therefore estimate the aspect ratio of the correlated structures within the perpendicular plane as
\beq
\frac{\lambda}{\xi}\sim \sin\theta.
\label{eq:anisalirelation}
\eeq
Combining \eqsand{eq:xi}{eq:anisalirelation} with \eqref{eq:tnl} gives us back 
the critical balance conjecture: 
\beq
\frac{\dzp\lpar}{\xi\vA} \sim \frac{\tA}{\tnl} \sim 1. 
\label{eq:CB}
\eeq

In combination with the parallel anisotropy, the above argument implies that the turbulence may be 3D anisotropic with respect to an instantaneous local basis defined by the directions of the mean magnetic field, the fluctuations, and the direction perpendicular to both. This was indeed confirmed in numerical simulations \citep{mallet3d,verdini2015} and in the solar wind \citep{chen3d}. Thus, the 3D anisotropy of RMHD turbulence can be understood as arising from a combination of critical balance \citep{gs95} and dynamic alignment \citep{boldyrev}. 

Another distinctive feature of RMHD turbulence (in which it resembles hydrodynamic turbulence; see, e.g. \citealp{frischbook}) is intermittency, i.e., the fact that the distribution of turbulent random fields is not scale-invariant (see \citealp{Chandran14} and references therein). It has become clear in recent years that intermittency is deeply intertwined with the physics of critical balance and dynamic alignment. For example, \cite{rcb} showed that, while nearly every random variable in numerically simulated RMHD turbulence is highly intermittent, the critical balance parameter $\chi \doteq \tA/\tnl$ has a distribution that is scale-invariant in the inertial range---as long as dynamic alignment is included in the definition of $\tnl$ as in \eqref{eq:tnl}. Moreover, it was shown in the same paper that the dynamic alignment angle was anticorrelated with amplitude at each given scale, i.e., the joint distribution of the turbulent random variables is highly non-trivial. \cite{mallet3d} then measured the intermittency of the turbulence in the local basis defined by the directions of the mean magnetic field, the fluctuations, and the direction perpendicular to both, and found the intermittency (quantified by structure-function scaling exponents) to be different in every direction. 

In view of this emerging evidence, it is essential to develop holistic theories that combine realistic treatments of critical balance and dynamic alignment with models for the intermittency of the turbulent fluctuations. Recently, a new theory of intermittent RMHD turbulence was proposed \citep{Chandran14}, which accurately predicted the scalings measured in the perpendicular direction by \cite{mallet3d}, by incorporating intermittency, critical balance and dynamic alignment into a physical model of the collisions of Alv\'enic structures. In this paper, we will take another approach and use these phenomena to propose a statistical model of the ``RMHD turbulent ensemble'', further constrained by assumptions about the geometrical structure of the turbulent fluctuations, and leading to prediction of the scalings in the perpendicular, parallel and fluctuation directions. We will begin by proposing a joint distribution of the relevant turbulent variables, and then fix all the parameters of this model by using physically motivated conjectures.

\section{RMHD ensemble}\label{sec:ensemble}

Suppose that we can meaningfully model the turbulent system as an ensemble of ``structures" or ``fluctuations", each defined by joint realizations of the following random variables: 
\beq
\begin{split}
\delta z &: \text{field amplitude}, \\
\lambda &: \text{perpendicular scale}, \\
\lpar &: \text{parallel scale},\\
\xi &: \text{fluctuation-direction scale}.
\end{split} \label{eq:rvs}
\eeq
We have made the significant simplification that we do not need two separate amplitudes $\delta z_\perp^\pm$;  i.e., we have restricted ourselves to the case of overall balanced turbulence, and assume that even locally, $\delta z_\perp^+ \sim \delta z_\perp^- \sim \delta z$. 

\subsection{Joint Probability Distribution}

Picking one particular structure from this ``RMHD turbulent ensemble'' corresponds to sampling the joint distribution $P(\delta z, \lambda,\lpar,\xi)$. Conditional structure functions, which can be measured in a real (or numerically simulated) turbulent system \citep{mallet3d}, correspond to moments of the conditional probability distributions $P(\delta z | \lambda)$, $P(\delta z | \lpar)$, $P(\delta z | \xi)$. We will therefore propose a particular functional form for the joint distribution $P(\delta z, \lambda,\lpar,\xi)$ of our model ensemble and then use this to calculate the conditional distributions in order to predict the scalings of the conditional structure functions. It will turn out that, with some additional assumptions, we can treat $\xi$ as a dependent variable, so we remove it from our consideration for now.

We will also need some global properties of the system, which we will treat as non-random:
\beq
\begin{split}
\epsilon &: \text{global cascade rate}, \\
v_{\rm A} &: \text{Alfv\'en speed}, \\
L_\perp &: \text{perpendicular outer scale}, \\
\Lpar &: \text{parallel outer scale}, \\
\overline{\delta z} &: \text{outer-scale fluctuation amplitude}. \\
\end{split}
\eeq
We stress that the assumption that these quantities are nonrandom---especially in the case of $\overline{\delta z}$---is a significant idealisation, but we believe this to be acceptable because none of them is scale-dependent.\footnote{In general, $\overline{\delta z}$ may be a scale-independent random variable, whose distribution is possibly non-universal and dependent on the details of the outer-scale energy injection. However, in Section \ref{sec:numpdf}, we shall see that treating it as non-random is at least not an outrageous idealisation: $\overline{\delta z}$ can be fit by a single constant, a few times the rms value of the Elsasser field.} To simplify further calculations, we define normalized random variables
\beq
\delta \hat{z} = \frac{\delta z}{\overline{\delta z}}, \quad \hlam = \frac{\lambda}{L_\perp}, \quad \hlpar=\frac{\lpar}{\Lpar}, \quad \hat{\xi} = \frac{\xi}{L_\perp}. 
\eeq

\begin{conj}
\emph{The fluctuation amplitudes can be modelled as
\beq
\delta \hat{z} = \beta^q,
\label{eq:betaq}
\eeq
where $q$ is a non-negative random integer, $\beta$ is a constant, $0\leq \beta \leq 1$, and the joint probability distribution of $q$, $\hlpar$ and $\hlam$ has density
\beq
P(q,\hlpar,\hlam) = \frac{\mu^q}{q!}e^{-\mu} f\lt(\frac{\hlpar}{\hlam^\alpha}\rt),\label{eq:jdist}
\eeq
where $\mu=\mu(\hlpar)$, $\alpha$ is a constant parameter and $f$ is some function.}
\end{conj}
The unknown function $f$ (which, it will turn out, we do not need to know, as long as it satisfies certain constraints) parametrises the anisotropy. We will examine the quality of this parametrisation in Section \ref{sec:anisdist}. The rest of the functional form of $P$ can be motivated by the following argument. 

\subsection{Log-Poisson Statistics}
\label{sec:poisson}

Let us calculate the conditional distribution $P(q | \hlpar)$ from the model \exref{eq:jdist} (this is a distribution of Elsasser-field increments across a fixed parallel point separation $\lpar$). Since $\mu$ is not a function of $\hlam$, we may integrate \eqref{eq:jdist} over $\hlam$ and obtain
\beq
P(q | \hlpar) = \frac{P(q,\hlpar)}{P(\hlpar)} = \frac{\int_0^1 f\lt(\hlpar/\hlam^\alpha\rt) d\hlam}{P(\hlpar)} \frac{\mu^q}{q!} e^{-\mu}. \label{eq:qglpar}
\eeq
Summing \eqref{eq:jdist} over $q$, we obtain
\beq
P\left(\hlpar,\hlam\right) = f\left(\frac{\hlpar}{\hlam^\alpha}\right),\label{eq:jdist2}
\eeq
whence, integrating over $\hlam$, we find 
\beq
P\lt(\hlpar\rt) = \int_0^1 f\lt(\frac{\hlpar}{\hlam^\alpha}\rt) d\hlam.\label{eq:lpardist}
\eeq
Therefore, from \eqref{eq:qglpar},
\beq
P(q | \hlpar) = \frac{\mu^q}{q!} e^{-\mu},\label{eq:qglpar2}
\eeq
which is a Poisson distribution with mean $\mu$.

Historically, the Poisson distribution as a model for the distribution of the logarithm of the fluctuation amplitude was used very successfully in hydrodynamic turbulence \citep{sl94,dubrulle94,shewaymire}. Moreover, there is recent direct observational evidence of the solar-wind turbulence being at least consistent with log-Poisson statistics \citep{zhdankin16}. The intermittency model of \cite{Chandran14}, which correctly predicts the perpendicular scalings of numerical RMHD turbulence, also used a log-Poisson model. An attractive physical interpretation of the log-Poisson model is that, as each fluctuation cascades to smaller $\lambda$ and $\lpar$, it undergoes an integer number $q$ of ``modulation defects", each of which is modelled as reducing the amplitude by a factor $\beta$ [see \eqref{eq:betaq}]. In \citet{Chandran14}, these defects were interpreted as collisions between unaligned Alfv\'enic wave packets.\footnote{The part of the cascade that does not involve the modulation defects involves sharpening in scale, which was linked by \citet{Chandran14} to collisions between aligned Alfv\'enic wave packets. They showed analytically that the amplitude of the fluctuations did not change in the collisions between aligned wave packets, which is why \eqref{eq:betaq} need not include a scale-dependent factor as the equivalent expression did in the hydrodynamic turbulence model of \citet{shewaymire}.} Whereas \cite{Chandran14} posited a log-Poisson distribution for fluctuation amplitudes $\delta z_\lambda$ conditioned on the perpendicular scale $\lambda$ (i.e., Elsasser-field increments across perpendicular point separations $\lambda$), we have here conjectured a log-Poisson distribution for $\delta z_{\lpar}$ conditioned on the parallel scale $\lpar$ (field increments across parallel separations $\lpar$). This can be justified in the following way.  

For the purposes of understanding intermittency, the constant-flux assumption \exref{eq:eps} can turned into a critically-balanced-RMHD version of the refined-similarity hypothesis \citep{k62}: 
\beq
\frac{\delta z^2_{\lpar}\vA}{\lpar} \sim \epsilon_{\lpar},
\label{eq:rsh}
\eeq
where $\epsilon_{\lpar}$ is the dissipation rate averaged over scale $\lpar$, but fluctuating 
over the entire box of length $\Lpar$. The global mean of this dissipation rate  
the Kolmogorov energy flux $\langle\epsilon_{\lpar}\rangle = \epsilon_{\Lpar} = \epsilon$, independent of scale. 
One might then argue, following \cite{k62}, that, refining the outer scale $\Lpar$ by  
a factor $a<1$, we must have 
\beq
\epsilon_{a\Lpar} = \epsilon_{\Lpar}W_1 = \epsilon W_1,
\eeq 
where $W_1$ is a positive random variable with $\langle W_1\rangle = 1$. Iterating this 
procedure, we may find that at any smaller scale $\lpar = a^k\Lpar$, 
\beq
\epsilon_{\lpar} = \epsilon \prod_{i=1}^{k} W_i, 
\eeq
where $W_i$'s are all independent
and identically distributed, with $\langle W_i\rangle = 1$. Since the distribution 
of $\epsilon_{\lpar}$ cannot depend on the (arbitrary) refinement constant $a$, 
we must be able to represent $\epsilon_{\lpar}$ as a product of an arbitrary number 
of these $W_i$'s and so the distribution of $\log\epsilon_{\lpar}$ must be infinitely divisible. 
Finally, since, by \eqref{eq:rsh}, $\epsilon_{\lpar} \propto \delta z^2_{\lpar}$, 
$\log \delta z_{\lpar}$ must also be infinitely divisible.\footnote{Note that a similar argument 
for $\delta z_\lambda$ is somewhat less straightforward because it would require constructing 
a refined similarity hypothesis starting from the first equality in \eqref{eq:eps} 
and letting $\tau_\mathrm{c}\sim\tnl$, with $\tnl$ given by \eqref{eq:tnl} 
\citep[as was done by ][]{Chandran14}. 
The resulting relationship between $\delta z_\lambda$ and $\epsilon_\lambda$ (whose logarithm 
is infinitely divisible by the same argument as explained above) 
then involves the alignment angle, whose distribution is {\em a priori} unknown. \label{fn:parvsperp}}  
The logarithms of the amplitudes are therefore described by a L\'evy process, which can always be written as the sum of a Gaussian process, a superposition of compound Poisson processes, and a non-random component \citep{satobook}. The Gaussian part is ruled out because it leads to a mathematically impossible scaling of structure functions \citep{frischbook}, and the simplest possible compound Poisson distribution is just a Poisson distribution. 

Thus, the log-Poisson model is at least reasonable physically, mathematically, and observationally (in Section \ref{sec:numpdf}, we will examine how well it fits numerical data). 
We now have an explicit joint distribution of the turbulent random variables, \eqref{eq:jdist}, which involves unknown parameters $\alpha$ and $\beta$ and unknown functions $\mu$ and $f$. In order to calculate the scalings of the conditional structure functions, these need to be determined or constrained using further physically motivated conjectures.

\section{Finding $\mu$ and $\alpha$: a turbulence of flux sheets}\label{sec:fluxsheets}
Taking some inspiration from \cite{sl94}, we consider the most intense structures in our 
turbulent ensemble (i.e., those with $q=0$) and assume that they are {\em not} space-filling. 
Based on observations of sheet-like structures in the solar wind \citep[e.g.,][]{greco09,perri12,osman14rec,chasapis15} and in numerically simulated turbulence, as well as theoretical considerations 
\citep[e.g.,][]{boldyrev,zhdankin13,zhdankin16intcy,Chandran14,howes2015}, 
we propose
\begin{conj}
\emph{The most intense structures are sheets transverse to the local perpendicular direction.}
\end{conj}
Mathematically, this means that, if we condition our distribution \exref{eq:jdist} on 
$\hlam$ (i.e., restrict ourselves to consider only field increments across both the mean- 
and the fluctuating-field directions), we must find that the filling fraction of the most 
intense structures is 
\beq
P(q=0 | \hlam) \propto \hlam.\label{eq:q0glam}
\eeq
We will find use for this requirement in Section \ref{sec:alpha}, but, since our distribution 
is formulated most compactly in terms of amplitudes conditional on $\lpar$, \eqref{eq:qglpar2}, 
the most accessible quantity for us is, in fact,  
\beq
P(q=0 | \hlpar) = e^{-\mu(\hlpar)}
\eeq
and so we can determine $\mu(\hlpar)$ if we can determine the filling fraction of 
our sheets as a function of $\hlpar$. Namely, anticipating 
\beq
P(q=0 | \hlpar) \propto \hlpar^\sigma,\label{eq:codim}
\eeq
we obtain
\beq
\mu(\hlpar) = -\sigma\ln\hlpar.\label{eq:musigma}
\eeq

\subsection{Finding $\sigma$: Refined Critical Balance}
\label{sec:sigma}

Consider what we expect the filling fraction of the singular sheets to be conditionally on $\hlpar$. By integrating out $\hlam$ [\eqref{eq:qglpar}], we have restricted consideration to field increments $\delta z$ between point separations that lie in a plane defined by the fluctuation direction and the parallel direction. This restricts us to a plane that is tangent to a ``flux sheet" that coherently extends a distance $\lpar$ along the mean-magnetic-field direction and a distance $\xi$ along the fluctuation direction (which, by definition, is perpendicular to both $\lpar$ and $\lambda$). Therefore the filling fraction of the sheet within the plane must be
\beq
P(q=0 | \hlpar) \sim \hlpar\hat{\xi}.\label{eq:q0glpar}
\eeq
Note that $\hat{\xi}$ is, by assumption, a function of both $q$ (i.e., the amplitude $\delta z$) and $\hlpar$ [see \eqref{eq:xi}], but we are about to argue that for $q=0$ it only depends on $\hlpar$. 

We now postulate the ``refined critical balance", a conjecture inspired by numerical evidence \citep{rcb}:
\begin{conj}
\emph{The ratio of linear to nonlinear time scales [see \eqref{eq:CB}]
\beq
\chi \doteq \frac{\delta z \lpar}{\xi\vA} = \lt(\frac{\Lpar}{L_\perp} \frac{\overline{\delta z}}{\vA}\rt) \beta^q \frac{\hlpar}{\hat{\xi}} \label{eq:chi}
\eeq
is statistically independent of scale.}
\end{conj}
This implies that for the most intense fluctuations, which have $q=0$, 
\beq
\hat{\xi} \sim \hlpar, 
\label{eq:xilpar}
\eeq
or, to be precise, the ratio $\hlpar / \hat{\xi}$ has a probability distribution that is statistically independent of scale. We therefore posit that \eqref{eq:q0glpar} becomes  
\beq
P(q=0 | \hlpar) \propto \hlpar^2.
\eeq
i.e., $\sigma=2$ in \eqref{eq:codim}. Consequently, from \eqref{eq:musigma}, 
\beq
\mu(\hlpar)= - 2\ln\hlpar.\label{eq:mu}
\eeq

Note that some circumstantial evidence in support of \eqref{eq:xilpar} has recently been reported by \cite{zhdankin16intcy}, who find that the lengths (our $\lpar$) and widths (our $\xi$) of Elsasser vorticity (current) structures in their numerical simulations have a joint distribution peaked at $\lpar\propto\xi$.

\subsection{Finding $\alpha$}
\label{sec:alpha}
To find $\alpha$, we will use \eqref{eq:q0glam}, and so 
we must first calculate 
\beq
P(q | \hlam) = \frac{P(q,\hlam)}{P(\hlam)}.\label{eq:cc}
\eeq
Integrating \eqref{eq:jdist2} over $\hlpar$, we obtain
\beq
P(\hlam) = \int_0^1f\left(\frac{\hlpar}{\hlam^\alpha}\right)d\hlpar = \hlam^{\alpha}\int_0^{1/\hlam^{\alpha}}f(y) dy.\label{eq:bb}
\eeq
Using \eqref{eq:mu} and integrating \eqref{eq:jdist} over $\hlpar$, we obtain
\begin{align}
P(q, \hlam) &= \int_0^1 \frac{[-2\ln\hlpar]^q}{q!} \hlpar^2 f\left(\frac{\hlpar}{\hlam^\alpha}\right)d\hlpar \nonumber \\
&= \hlam^{{3}{\alpha}}\int_0^{1/\hlam^{{\alpha}}}\frac{[-2\ln (y \hlam^{{\alpha}})]^q}{q!} y^2 f(y) dy.\label{eq:aa}
\end{align}
With \eqsand{eq:aa}{eq:bb}, \eqref{eq:cc} gives us
\beq
P(q | \hlam) = \frac{\hlam^{{2}{\alpha}}}{q!}\frac{\int_0^{1/\hlam^{{\alpha}}}[-2\ln (y \hlam^{{\alpha}})]^q y^2 f(y) dy}{\int_0^{1/\hlam^{\alpha}}f(y) dy}.\label{eq:qglamfull}
\eeq
This is a weighted mixture of Poisson distributions with different means. 
The probability (filling fraction) of the most intense structures conditional on $\hlam$ is, 
therefore,
\beq
P(q=0 | \hlam) = \hlam^{2{\alpha}}\frac{\int_0^{1/\hlam^{{\alpha}}} y^2 f(y) dy}{\int_0^{1/\hlam^{{\alpha}}}f(y) dy}.
\label{eq:Pq0lam}
\eeq
Suppose that $f(y)$ decays fast enough so, at most, 
\beq
f(y) = O\left(\frac{1}{y^{\delta +3}}\right)\text{ for } \delta > 0\text{ as }y \to \infty.\label{eq:ordercond}
\eeq
Then, in the inertial range, i.e., for $\hlam\ll1$ (in the limit of large Reynolds numbers), \eqref{eq:qglamfull} becomes
\beq
P(q=0 | \hlam) \approx C \hlam^{2{\alpha}},\label{eq:qglam3}
\eeq
where 
\beq
C = \frac{\int_0^\infty y^2 f(y) dy}{\int_0^\infty f(y) dy} = \mathrm{const}. 
\eeq
Comparing \eqsand{eq:qglam3}{eq:q0glam}, we conclude that 
\beq
\alpha=\frac{1}{2}.\label{eq:alpha}
\eeq
Besides enabling us to fix the parameters of our model \exref{eq:jdist}, this result will have 
interesting and checkable consequences, which will be examined in Section \ref{sec:anisdist}.\\ 

We now have a complete expression for our joint probability distribution \exref{eq:jdist}:
\beq
P(q,\hlpar,\hlam) = \frac{(-2\ln\hlpar)^q}{q!}\hlpar^2 f\lt(\frac{\hlpar}{\hlam^{1/2}}\rt).
\eeq
It remains to determine the parameter $\beta$, which determines the relationship 
between $q$ and $\delta\hat{z}$ [\eqref{eq:betaq}], and hence calculate all desired 
scalings.  

\section{Scalings}

\subsection{Finding $\beta$: Parallel Cascade}
\label{sec:beta}
To find $\beta$, we use the constant-flux hypothesis \exref{eq:eps}, 
combined with the critical-balance hypothesis and formalised as follows:
\begin{conj}
\emph{The mean flux of energy is constant through parallel scales in the inertial range:
\beq
\epsilon = \frac{\langle \delta z^2 | \lpar \rangle \vA}{\lpar} = \rm{const}.\label{eq:conjeps}
\eeq
}
\end{conj}

Let us calculate $\langle \delta \hat{z}^n | \hlpar \rangle$ by multiplying \eqref{eq:qglpar2} by $\delta \hat{z}^n= \beta^{nq}$, summing over $q$, and using \eqref{eq:musigma}:
\beq
\langle \delta \hat{z}^n | \hlpar \rangle = \sum_{q=0}^\infty \beta^{nq} \frac{\mu^q}{q!} e^{-\mu}
= e^{-\mu(1-\beta^n)} = \hlpar^{\sigma(1-\beta^n)}.\label{eq:parsf}
\eeq
Fitting the case of $n=2$ to \eqref{eq:conjeps}, we obtain a simple equation for $\beta$,
\beq
\sigma (1-\beta^2) = 1,
\eeq
whose positive solution for $\sigma=2$ [\eqref{eq:mu}] is
\beq
\beta=\frac{1}{\sqrt{2}}.\label{eq:beta}
\eeq
Thus, all the parameters of our model have now been determined.

\begin{figure*}
\begin{tabular}{cc}
\includegraphics[width=9cm]{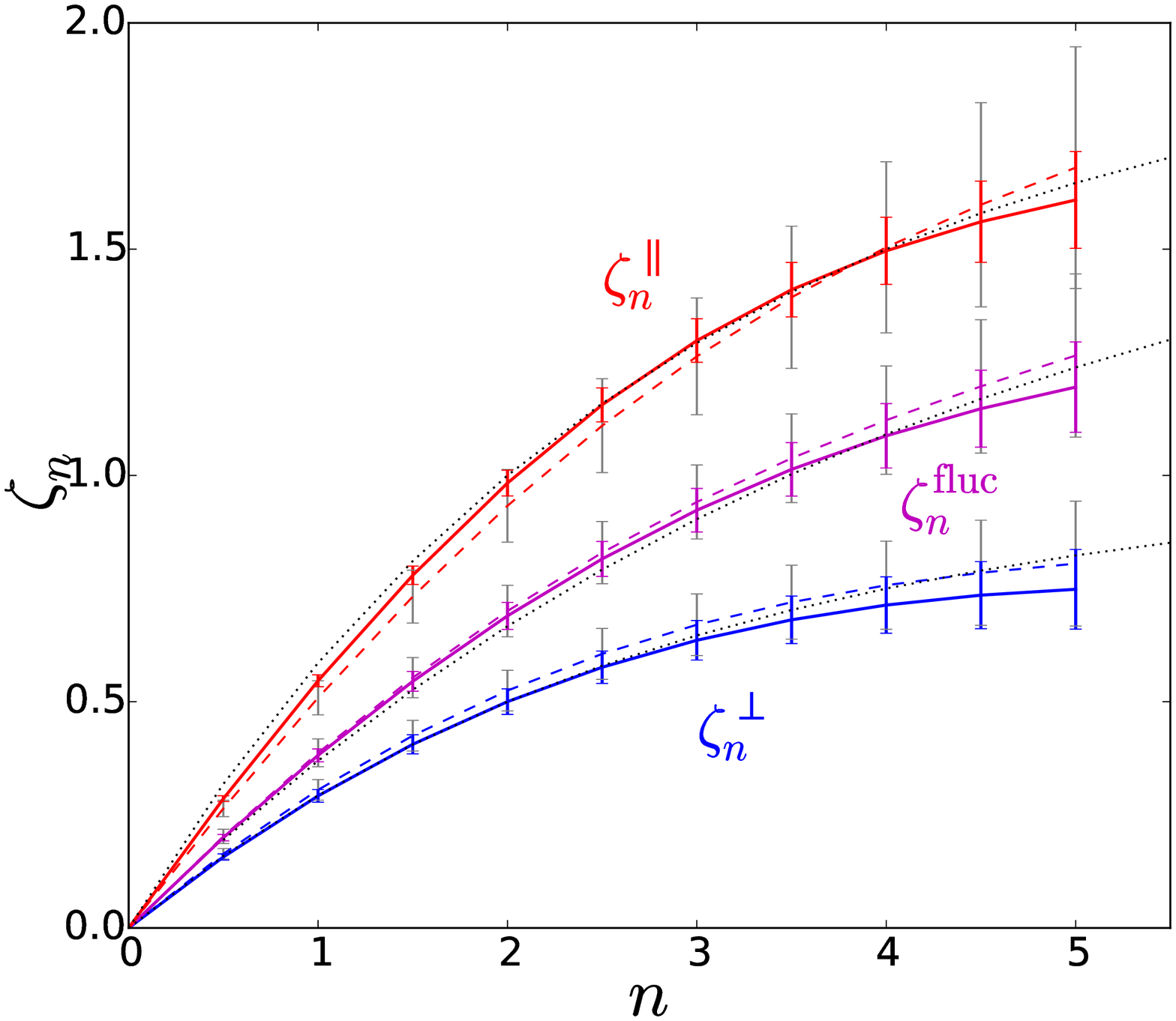} &
\includegraphics[width=9cm]{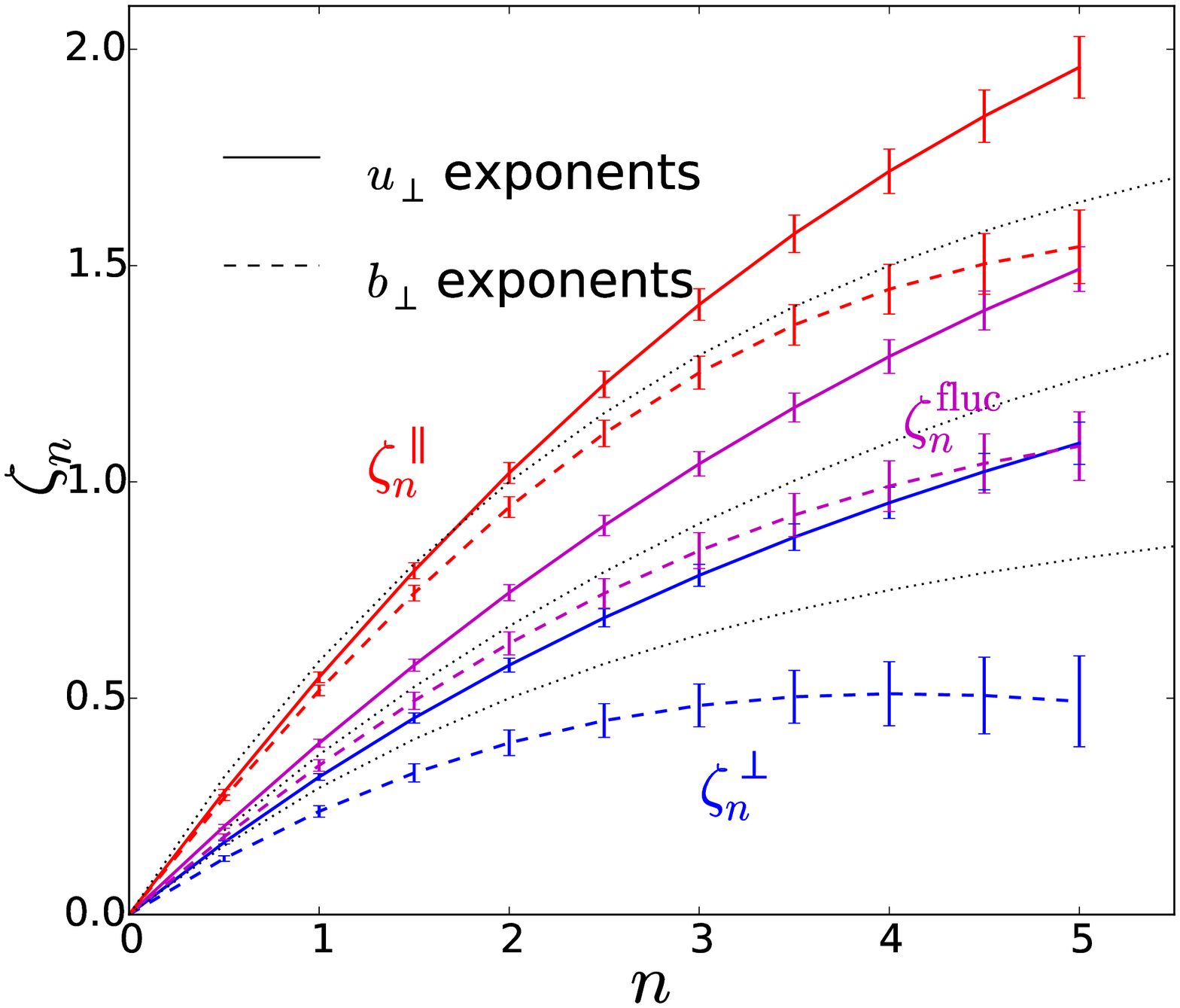} \\
\textbf{(a)} & \textbf{(b)} \\
\end{tabular}
\caption{\textbf{(a)} Scaling exponents of the $n$th-order structure functions of $\vzp^+$ calculated from a $1024^3$ RMHD numerical simulation of \citet{mallet3d} are shown as solid lines (see their Figure 1B, showing how the scaling exponents were fitted). They correspond to point separations within $10\degree$ of the perpendicular (``$\perp$'', blue), fluctuation (``$\rm fluc$'', purple) and parallel (``$\parallel$'', red) directions ($10\degree$ in the context of the parallel direction refer to angles calculated using lengths in code units; see footnote \ref{fn:lengths}). The dotted lines are theoretical predictions given by \eqsref{eq:zetaperp}, \exref{eq:zetapar} and \exref{eq:zetafluc}. Error bars are standard deviations calculated from data from 10 snapshots separated by more than a turnover time. For further details, see \citet{mallet3d}. To illustrate the level of numerical convergence (or otherwise) of these results, we also show (as dashed lines with grey error bars) the scaling exponents obtained from a smaller, $512^3$, but otherwise identical, simulation.
\textbf{(b)} Scaling exponents for velocity $\mathbf{u}_\perp$ (solid lines) and magnetic field $\mathbf{b}_\perp$ (dashed lines) for the same $1024^3$ simulation. The fluctuation direction in each case is aligned with the field for which the exponents are calculated. See the discussion in Section \ref{sec:numsf}.
}
\label{fig:3dsfscalings}
\end{figure*}

\subsection{Parallel Structure Functions}
In \eqref{eq:parsf}, we already calculated the scaling exponents of the parallel conditional structure functions: 
defining $\zeta_n^\parallel$ by 
\beq
\langle \delta z^n | \lpar \rangle \propto \lpar^{\zeta_n^{\parallel}},\label{eq:sfpar}
\eeq
we find 
\beq
\zeta_n^\parallel = \sigma (1-\beta^n) = 2\left(1- \frac{1}{2^{n/2}}\right) .\label{eq:zetapar}
\eeq
The second-order exponent is 
\beq
\zeta_2^\parallel = 1\label{eq:zetapar2}
\eeq
(by assumption; see Section \ref{sec:beta}), implying the parallel spectral index of $-2$ 
\citep{gs97}. 

\subsection{Perpendicular Structure Functions}

To find the scaling exponents $\zeta_n^\perp$ of the perpendicular structure functions,
\beq
\langle \delta z^n | \lambda \rangle \propto \lambda^{\zeta_n^\perp},\label{eq:sfperp}
\eeq
we multiply \eqref{eq:qglamfull} by $\delta \hat{z}^n = \beta^{nq}$, and sum over $q$:
\beq
\begin{split}
\langle \delta \hat{z}^n | \hlam \rangle &= \frac{\hlam^{{2}{\alpha}}\int_0^{1/\hlam^{{\alpha}}}\sum_{q=0}^\infty\frac{[-2\beta^n \ln (y \hlam^{{\alpha}})]^q}{q!} y^2 f(y) dy}{\int_0^{1/\hlam^{\alpha}}f(y) dy}, \\
&= \hlam^{2 \alpha(1-\beta^n)}
\frac{\int_0^{1/\hlam^{{\alpha}}} y^{2(1-\beta^n)} f(y) dy}{\int_0^{1/\hlam^{{\alpha}}} f(y) dy} \\
&\approx C_n \hlam^{2 \alpha(1-\beta^n)}. 
\end{split}
\label{eq:ff}
\eeq
The last equality holds in the inertial range, i.e., in the limit $\hlam\ll1$, with 
\beq
C_n = \frac{\int_0^\infty y^{2(1-\beta^n)} f(y) dy}{\int_0^\infty f(y) dy} = \mathrm{const}.
\eeq
Since $\beta=1/\sqrt{2}<1$, the integrals converge provided that the condition \exref{eq:ordercond} 
holds. Finally, using $\alpha = 1/2$ [\eqref{eq:alpha}], we have 
the perpendicular scaling exponents: 
\beq
\zeta_n^\perp = 2\alpha(1-\beta^n) = 1- \frac{1}{2^{n/2}}. \label{eq:zetaperp}
\eeq
The second-order exponent is 
\beq
\zeta_2^\perp = \frac{1}{2},\label{eq:zetaperp2}
\eeq
implying the perpendicular spectral index of $-3/2$ \citep[cf.][]{boldyrev}.  

\subsection{Fluctuation-Direction Structure Functions}

The scalings of the structure functions conditional on $\xi$, 
\beq
\langle{\delta z^n | {\xi}}\rangle \propto \xi^{\zeta^{\rm fluc}_n},\label{eq:sfxi}
\eeq
are harder to determine because $\xi$ depends on the amplitude $\delta z$ as well as on $\lpar$ [see \eqref{eq:xi}]. Rather than taking this into account rigorously,\footnote{Which can perhaps be done via \eqref{eq:chi}, but leads to unilluminating and ultimately unrewarding calculations.} we will employ a simple ruse.  

Let us assume that the fluctuations that provide the dominant contribution to the $n$th-order structure function conditional on $\lpar$ are also those that provide the dominant contribution to the structure function conditional on $\xi$. Let $\delta z_{\mathrm{eff},n}$ be the amplitude of these fluctuations, namely, by definition,
\beq
\delta z_{\mathrm{eff},n} = \langle \delta z^n | {l_\parallel}\rangle^{1/n} \propto \lpar^{\zeta_n^\parallel/n}.
\label{eq:zeff} 
\eeq
Motivated by \eqref{eq:chi}, we now posit that these fluctuations have scale $\xi_{\mathrm{eff},n}$ in the fluctuation direction, given by 
\beq
\xi_{\mathrm{eff},n} = \frac{\delta z_{\mathrm{eff},n}\lpar }{\vA} \propto \lpar^{1+\zeta_n^\parallel/n}.
\eeq
Then, from \eqref{eq:zeff}, 
\beq
\lt(\delta z_{\mathrm{eff},n}\rt)^n \propto \xi_{\mathrm{eff},n}^{\zeta^{\rm fluc, eff}_n},
\eeq
where, using \eqref{eq:zetapar},  
\beq
\zeta^{\rm fluc, eff}_n = \frac{\zeta_n^\parallel}{1 + \zeta_n^\parallel/n}
= \frac{n(1-\beta^n)}{n/\sigma+1-\beta^n} 
= \frac{n\left(1-1/2^{n/2}\right)}{n/2+1-1/2^{n/2}}\label{eq:zetafluc}.
\eeq
For lack of a more quantitative theory, we will consider these to be an acceptable approximation of the exponents $\zeta^{\rm fluc}_n$ defined by \eqref{eq:sfxi}. The second-order exponent is 
\beq
\zeta^{\rm fluc, eff}_2 = \frac{2}{3}, \label{eq:zetafluc2}
\eeq 
implying the fluctuation-direction spectral index of $-5/3$ \citep[cf.][]{boldyrev}. Note also that as $n\to\infty$, 
\beq
\zeta^{\rm fluc, eff}_n \approx \zeta_n^\parallel \to 2
\quad\text{as}\quad n\to\infty. 
\eeq
This is in line with the idea that $\xi\propto\lpar$ for the most intense structures, which dominate the structure function as $n\to\infty$ (see Section \ref{sec:sigma}). 
Thus, while \eqref{eq:zetafluc} is not much more than a useful mnemonic, it behaves in a physically transparent way and, as we are about to see, also works quite well, so we consider it worthwhile, even if a more sophisticated theory is undoubtedly conceivable.

\begin{figure*}
\vskipfig
\includegraphics[width=5.8cm]{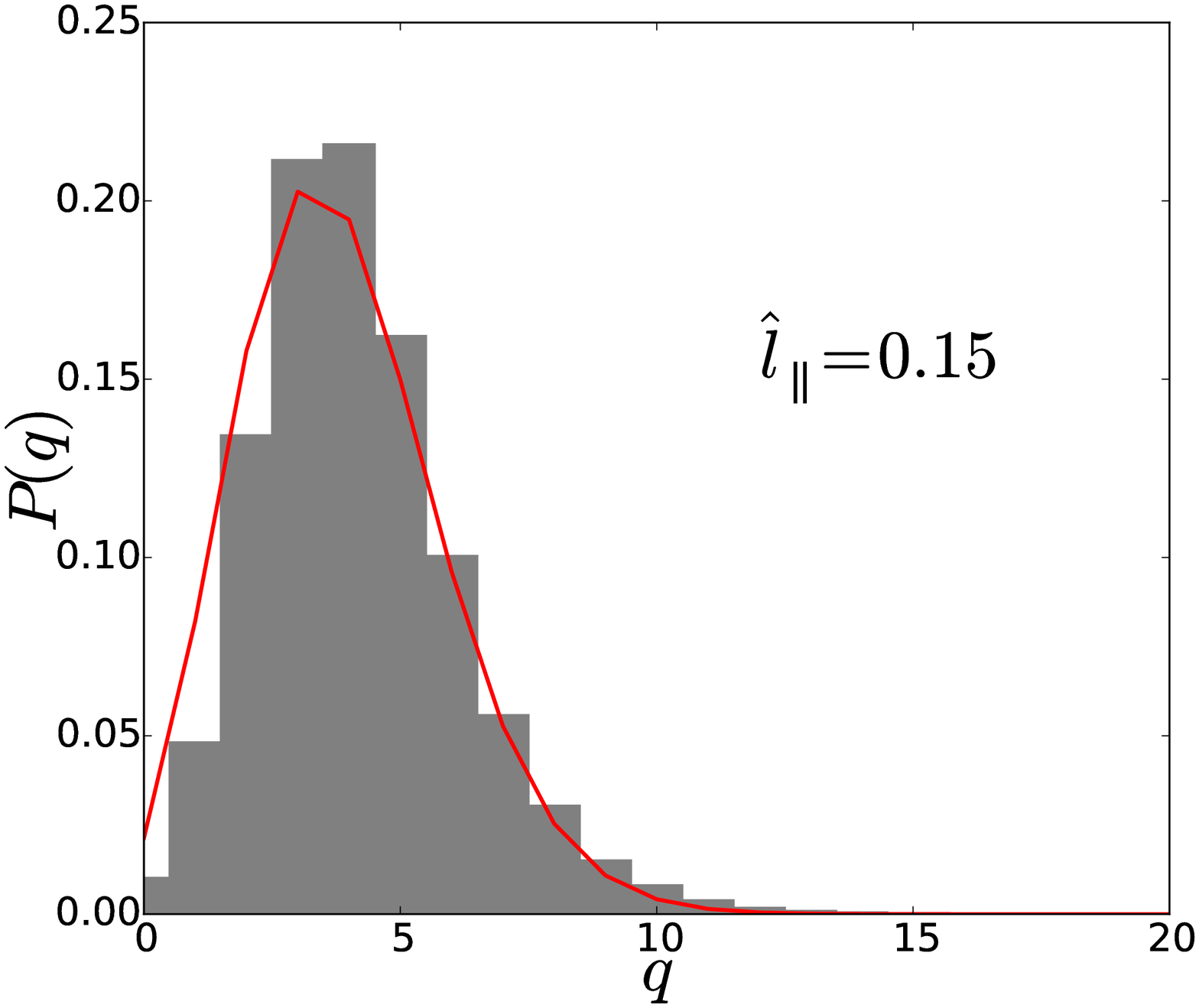}
\includegraphics[width=5.8cm]{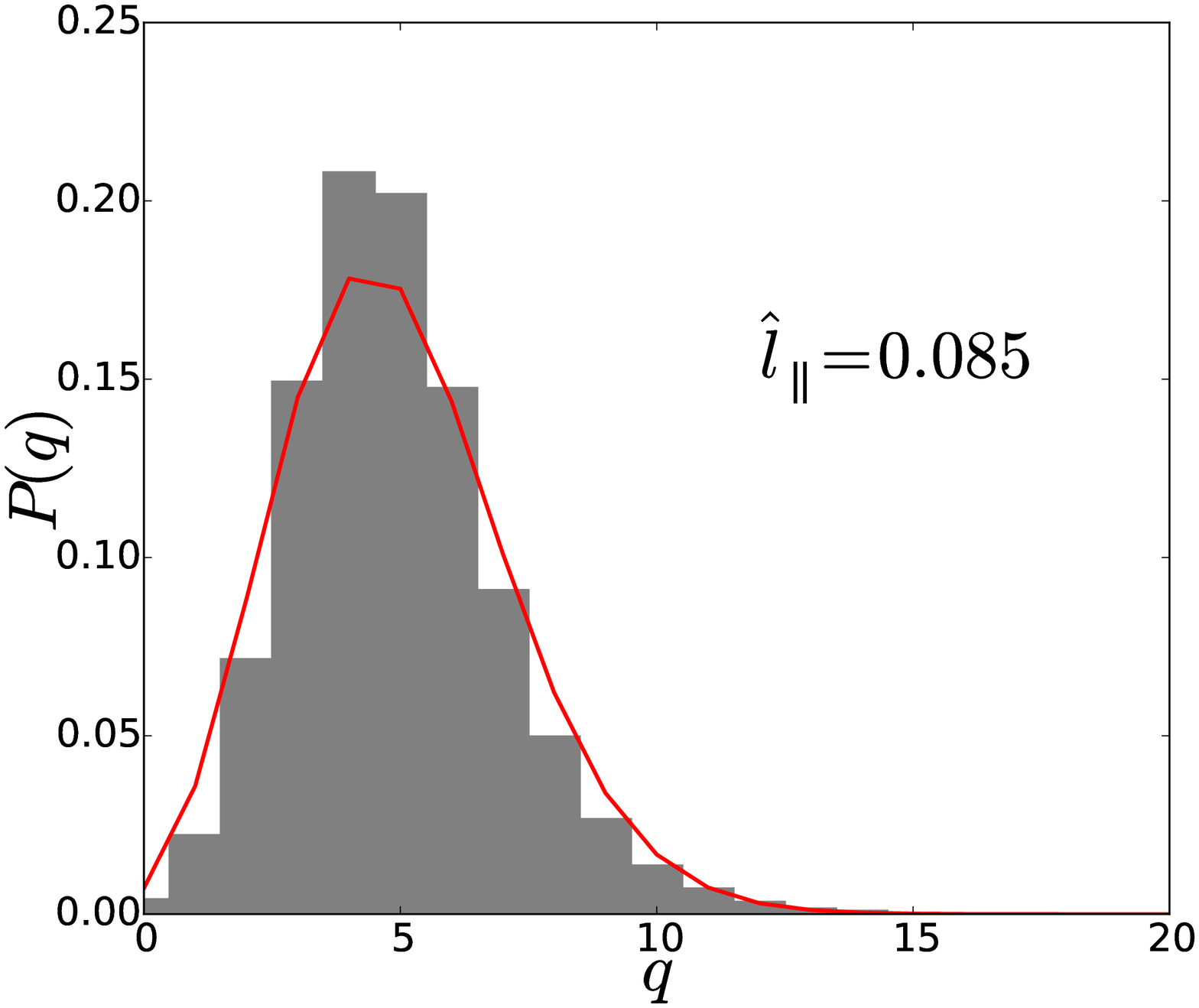}
\includegraphics[width=5.8cm]{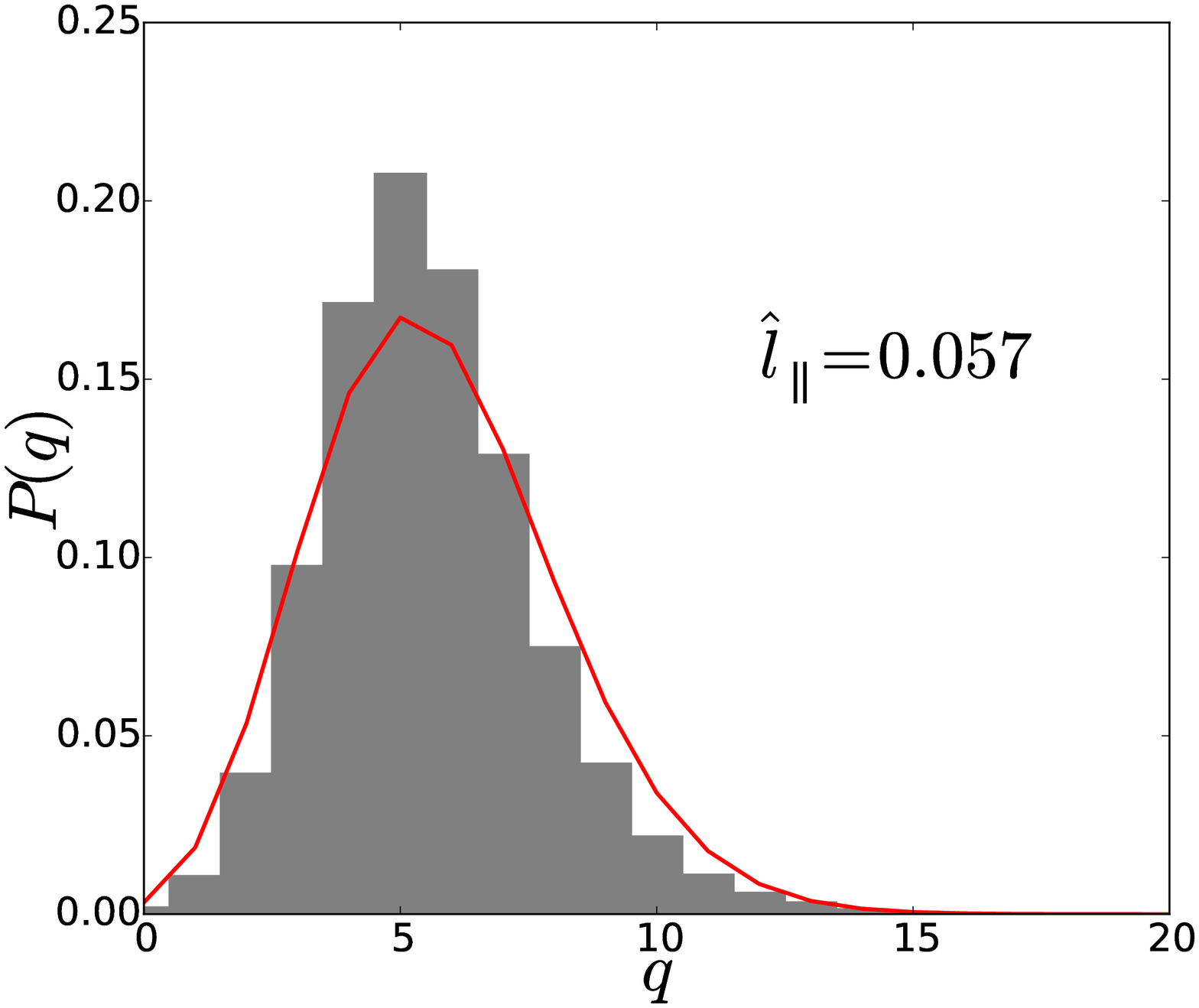}\\
\includegraphics[width=5.8cm]{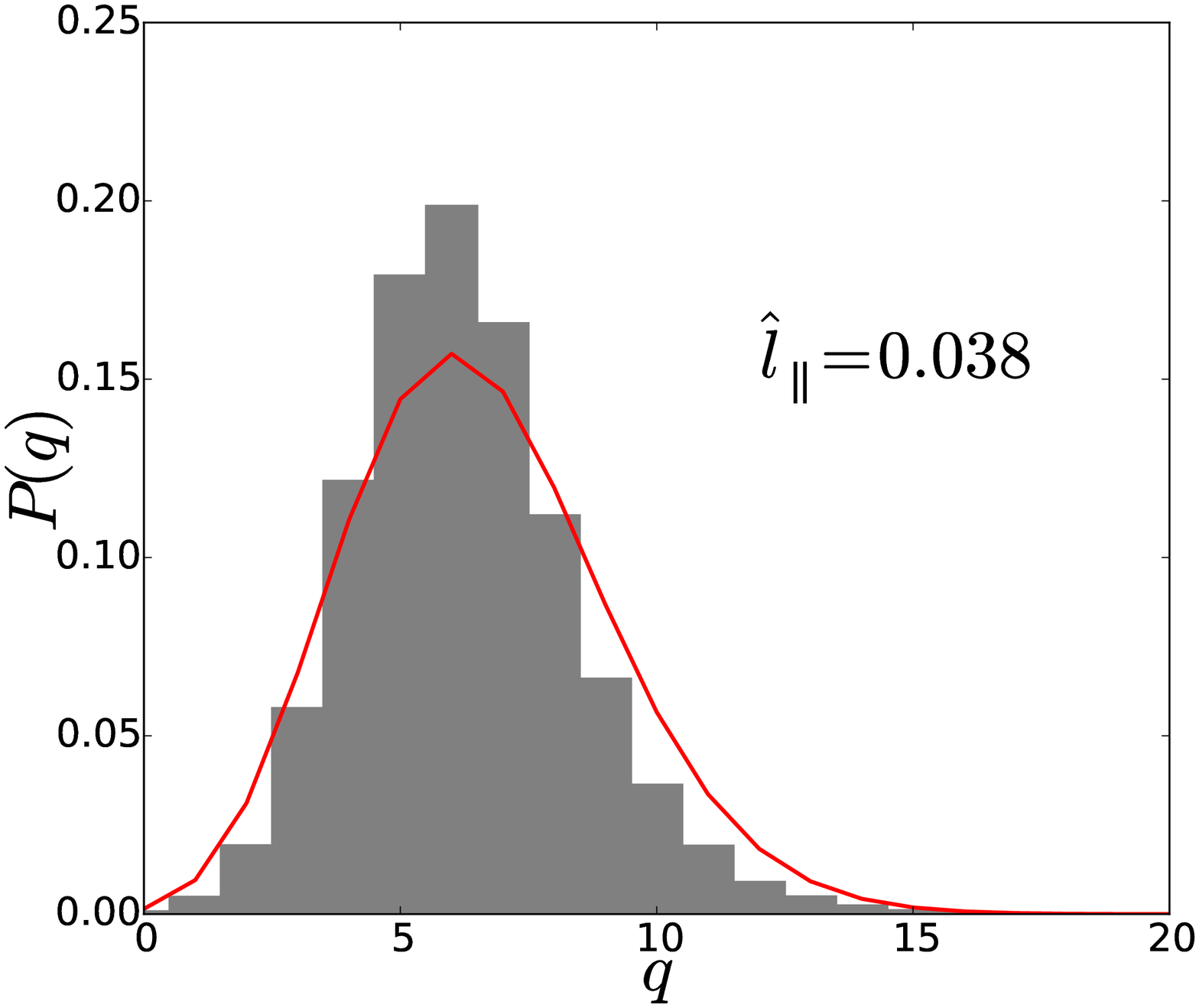}
\includegraphics[width=5.8cm]{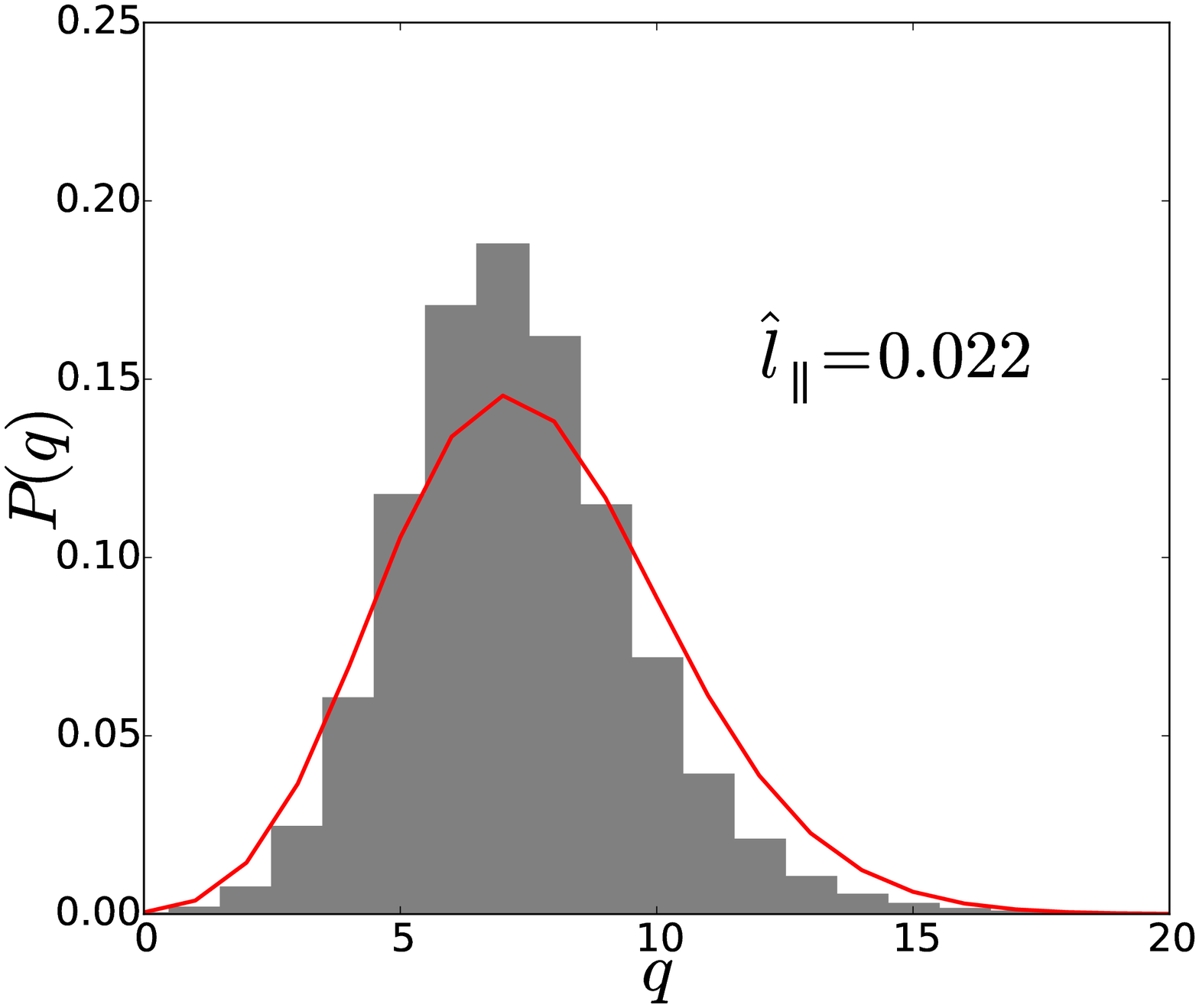}
\includegraphics[width=5.8cm]{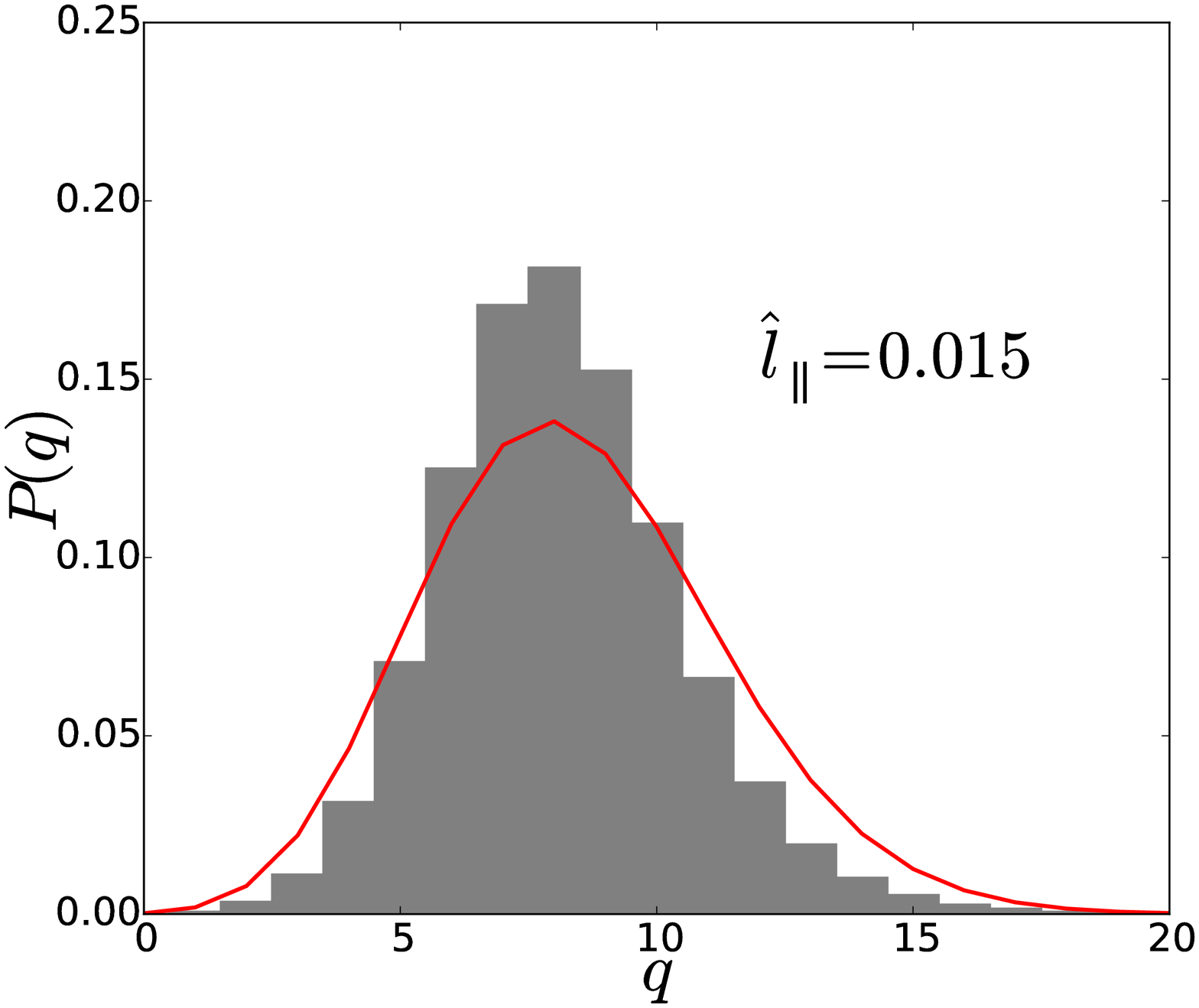}
\caption{Probability distributions of the random variable $q$ defined in \eqref{eq:defq}, calculated from the same simulation as in \figref{fig:3dsfscalings}, are shown as grey bars for six scales: $\hlpar = 0.15, 0.085, 0.057, 0.038, 0.022, 0.015$; $\hlpar$ is normalised to the parallel box size ($\Lpar=2\pi$ in code units). These scales approximately cover the inertial range (the plots of the structure functions vs.\ scale can be found in \citealt{mallet3d}). Red solid lines show the Poisson distribution \exref{eq:qglpar2} with mean $\langle q\rangle = \mu = -2\ln\hlpar$ [\eqref{eq:mu}] for those values of $\hlpar$. Based on \eqref{eq:zbarfit} applied to 17 logarithmically spaced scales (six of which are shown in this Figure) in the inertial interval $\hlpar\in[0.015,0.15]$, the best-fitting value $\overline{\delta z} = 9.68$ was found (in code units; in the same units, the rms value of the Elsasser field was $\langle|\vzp^+|^2\rangle^{1/2}=3.32$). Note that, while no special measures were taken to ensure that $q\ge0$, there were virtually no increments with $q<0$.\label{fig:poispar}}
\vskipfig
\end{figure*}

\section{Numerical Tests}
\label{sec:num}

\subsection{Structure-Function Scaling Exponents}
\label{sec:numsf}

To evaluate how well our model can describe available data, we compare the scaling exponents given by \eqsref{eq:zetapar}, \exref{eq:zetaperp} and \exref{eq:zetafluc} to those measured using conditional structure functions calculated for the $1024^3$ RMHD numerical simulation described in detail in \cite{mallet3d}. The scaling exponents measured in the simulation are reproduced in \figref{fig:3dsfscalings}{a} and show very reasonable agreement with our model. We refer the reader to \cite{Chandran14} for a review and discussion of earlier numerical 
and observational measurements of the structure-function exponents. Note in particular that 
results obtained in full-MHD \citep[e.g.,][]{mullerbiskamp}, rather than RMHD simulations, do not appear to be converged with respect to the asymptotically 
large size of the mean field, which in RMHD is analytically hard-wired by the underlying ordering. As the size of the mean field was increased, their measured high $n$ scaling exponents decrease, towards those of our model.

In \figref{fig:3dsfscalings}{b}, we give the scaling exponents for the velocity and magnetic fields (to contrast them with those for the Elsasser fields in \figref{fig:3dsfscalings}{a}). These do not coincide with either each other (${\bf b}_\perp$ is ``more intermittent'' than ${\bf u}_\perp$) or with the scalings for the Elsasser fields and are not well described by our Eqs. (\ref{eq:zetapar}), (\ref{eq:zetaperp}) and (\ref{eq:zetafluc}). This is not a particular problem for our theory, which does not claim to be able to predict these scalings---indeed, 
to do this, we would have had to construct a model for the {\em joint}
distribution of $\delta z^+_\perp$ and $\delta z^-_\perp$
within any given fluctuation (i.e., a statistical model of "local imbalance"
in RMHD turbulence). Presumably, the fact that the
velocity and magnetic-field perturbations have different scaling
properties than the Elssasser fields means that all these fields
cannot simply be assumed to be aligned with each other with
alignment angles that have similar scale dependence
\citep[cf.][]{pbxhel}. It is indeed a known property
of numerical MHD turbulence that alignment angles between different
fields can differ \citep{bl06,beresnyak09,mcatbolalign}. They can also differ in the MHD fluctuations measured in space
\citep{wicksalign,wicks13}. Some theoretical predictions of various
alignment angles can be found in \citet{Chandran14}. 

\subsection{Distribution of Parallel Increments}
\label{sec:numpdf}

Let us now attempt a more sensitive test and check whether the distribution of the fluctuation amplitudes conditional on the parallel scale is consistent with our log-Poisson model \exref{eq:qglpar2}. We do this by directly calculating the distribution of the {\em parallel} increments of the Elsasser field produced by our numerical simulation. Namely, 
in view of \eqref{eq:betaq}, we consider the random variable 
\beq
q=\frac{\ln\lt(\delta z^+_{\lpar}/\overline{\delta z}\rt)}{\ln\beta},
\label{eq:defq}
\eeq
where $\beta=1/\sqrt{2}$, as per \eqref{eq:beta}, and
$\delta z^+_{\lpar}$ is a field increment across point separation $\lpar$ within $10\degree$ of the direction parallel to the local mean field.\footnote{Angles are calculated formally using lengths in code units. In theory, in RMHD, the box is infinitely elongated and so the parallel units of length are arbitrarily rescalable with respect to the perpendicular ones as long as $v_A$ is rescaled by the same factor. In the code units, our box is cubic, with $L_\perp=\Lpar = 2\pi$. The ``local mean field'' is defined in the same way as in \eqref{eq:lpar}. \label{fn:lengths}} We treat $\overline{\delta z}$ as a {\em scale-independent} fitting parameter, determined by a least-squares linear fit between the mean of the distribution of $\ln\delta z^+_{\lpar}$, and the mean expected from our model, $\langle q\rangle = \mu$, or, using \eqref{eq:mu}, 
\beq
\lt\langle \ln\delta z^+_{\lpar}\rt\rangle = \ln\overline{\delta z} - 2\lt(\ln\hlpar\rt)\ln\beta.
\label{eq:zbarfit}
\eeq
Here we have na\"ively taken $\Lpar=2\pi$, the size of the box in code units (which is the forcing scale in our simulation), and normalised $\hlpar=\lpar/\Lpar$. The fit is done for a number of values of $\hlpar$, covering the extent of the inertial range in the simulation. Thus, we are fitting an entire family of scale-dependent distributions using a single scale-independent parameter, so finding them at least consistent with our log-Poisson model \exref{eq:qglpar2} would be a nontrivial and encouraging result.  

This is indeed the result that we find: the numerically computed distributions for several values of $\hlpar$ from our inertial interval are shown in \figref{fig:poispar}, superimposed on the theoretical curves tracing the model distribution \exref{eq:qglpar2} with its mean $\mu$ given by \eqref{eq:mu}. The agreement is reasonable, especially with regards to the position of the mean. The high-$q$ tails of the distributions agree slightly less well, which we believe to be due to a systematic underrepresentation of the high-$q$ structures with the two-point field increments\footnote{We note that two-point increments of a continuous Elsasser field are not 
strictly the same thing as amplitudes of notional individual "structures" or 
"fluctuations" of which our model "RMHD ensemble" consists (see Section \ref{sec:ensemble}). 
We do, however, use two-point increments as the most convenient and 
simple way to probe fluctuations at a particular scale and implicitly assume 
that these should have the same statistics. Clearly, such a correspondence
can only be approximate.}, as the contribution to the total field increment due to these structures can be small compared to the contribution from a Taylor expansion of the fields associated with structures at larger scales but having lower $q$. Another source of errors may be an insufficiently precise identification of the direction parallel to the local mean field. Note at any rate that high values of $q$ do not contribute strongly to the (large-$n$) structure functions because they correspond to {\em lower} amplitudes. This is presumably why the scaling exponents in the numerical simulation (\figref{fig:3dsfscalings}{a}) are captured so well by our model. 

We stress that the log-Poisson fit that we have obtained for $\delta z_{\lpar}$ is quite good, compared, for example, to the outcome of a similar procedure attempting to fit {\em perpendicular} increments $\delta z_\lambda$ to a log-Poisson model, as carried out, e.g., by \citet{zhdankin16}: they point out that, whereas the log-Poisson model for structure-function scaling exponents works well, the distribution of the field increments itself is not well fit by a log-Poisson curve; we find the same in our own numerical simulation. In our model, however, this is not a problem because the distribution of the logarithms of the perpendicular field increments, \eqref{eq:qglamfull}, is a Poisson mixture rather than a pure Poisson distribution. This might be viewed as a piece of circumstantial evidence in support of our argument (in Section \ref{sec:poisson}) that the parallel field increments are, via the constant-flux and critical-balance assumptions, more directly related to the infinitely divisible (and, therefore, likely log-Poisson) dissipation field than the perpendicular ones (see footnote \ref{fn:parvsperp}). 

\section{Distribution of Anisotropy}\label{sec:anisdist}

The joint distribution of $\lpar$ and $\lambda$ is given by \eqref{eq:jdist2}. 
It characterised the scale-dependent anisotropy of the turbulent structures in the RMHD ensemble. 
Using \eqref{eq:bb}, we find that, in the inertial range (where $\hlam\ll1$),  
\beq
P(\hlpar | \hlam) = \frac{P(\hlpar,\hlam)}{P(\hlam)} \approx 
\frac{f(\hlpar/\hlam^\alpha)}{\hlam^\alpha I},
\eeq
where $I = \int_0^\infty f(y) dy$, which we assume converges. 
Changing variables from $\hlpar$ to $y=\hlpar/\hlam^\alpha$, we find 
\beq
P(y | \hlam) = \frac{f(y)}{I}.\label{pyglam}
\eeq
Thus, our (thus far unknown) function $f$ is just the probability density function of $y$,  
which is independent of $\hlam$ (i.e., $y$ has scale-invariant statistics). 
This means that, for field increments deep enough into the inertial range,  
\beq
\hlpar \sim \hlam^\alpha. \label{eq:paraniseq}
\eeq
This is, of course, also why our \eqsand{eq:zetaperp}{eq:zetapar} had the property 
\beq
\zeta_n^\parallel = 2 \zeta_n^\perp,
\eeq
which we showed in \figref{fig:3dsfscalings} to be approximately true in numerically 
simulated RMHD turbulence. 
 
\begin{figure}
\includegraphics[width=8.9cm]{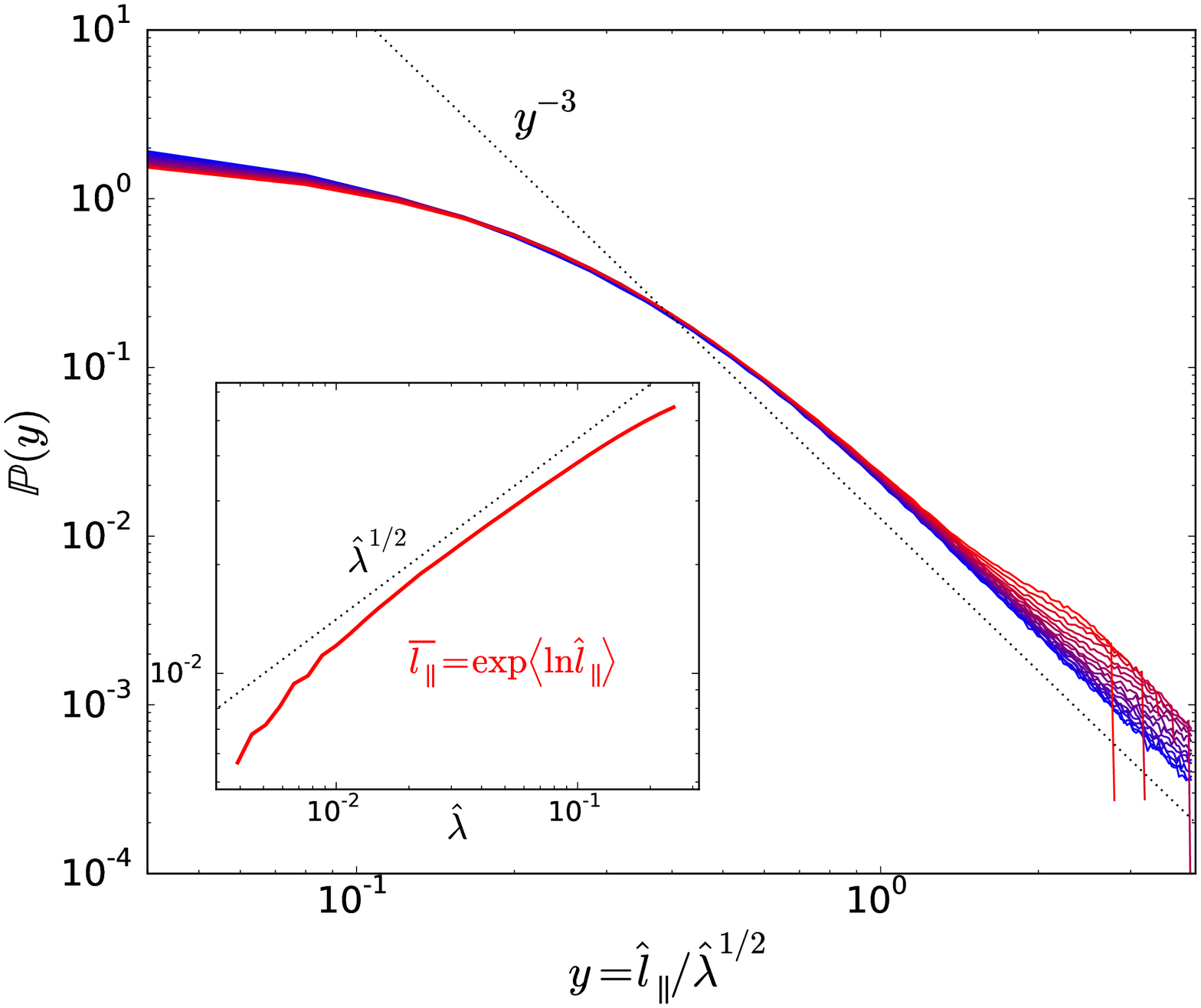}
\caption{The distribution of $y=\hlpar/\hlam^{1/2}$ for $\vzp^+$, calculated in the same simulation as in \figref{fig:3dsfscalings}. Here $\hlpar$ and $\hlam$ are normalised to box size in the parallel ($\Lpar=2\pi$ in code units) and perpendicular ($L_\perp=2\pi$ in code units) directions, respectively (which were also the scales at which the turbulence was forced; see \citealt{mallet3d}).The range of perpendicular scales is from $\hlam=0.015$ (blue/dark lines) to $\hlam=0.15$ (red/light lines). The dotted line shows the critical scaling $f\propto y^{-3}$ [see condition \exref{eq:ordercond}]. Inset: the ``typical" parallel scale obtained via logarithmic average $\overline{\lpar} = \exp \lt\langle\ln \hlpar \rt\rangle$; the $\overline{\lpar}\propto\lambda^{1/2}$ scaling is shown by the dotted line.\label{fig:lpardist}}
\end{figure}

Let us now measure the distribution of the anisotropy directly. We follow \cite{rcb}, who defined the parallel coherence length $\lpar$ for a given perpendicular increment $\lambda$ between spatial positions $\mathbf{r}_0$ and $\mathbf{r}_0+\mathbf{r}_\perp$ (where $|\mathbf{r}_\perp|=\lambda$) as the distance along the {\em perturbed} field line at which the Elsasser-field increment is the same as the perpendicular increment \citep{chovishniac,marongoldreich,Matthaeus12}:  
\begin{align}
\nonumber
&\lt|\vzp^\pm\lt(\vr_0 + \frac{\vrp + \lpar\vbloc}{2}\rt) - 
\vzp^\pm\lt(\vr_0 + \frac{\vrp - \lpar\vbloc}{2}\rt)\rt|\\
&= |\vzp^\pm(\vr_0+\vrp) - \vzp^\pm(\vr_0)|,
\label{eq:lpar}
\end{align}
where $\vbloc=\vBloc/|\vBloc|$ is the unit vector along the 
``local mean field'' $\vBloc\doteq\vB_0 + [\vbp(\vr_0)+\vbp(\vr_0+\vrp)]/2$. 
Using this definition, we can measure the distribution of $\lpar$ as a random variable 
conditional on $\lambda$. The resulting rescaled distribution of $y = \lpar/\lambda^{1/2}$ is shown in \figref{fig:lpardist}. Both the core of the distribution and the ``typical" values of $\lpar$ (defined in terms of a logarithmic average) appear to support the corollaries of our model that $y$ is scale invariant and $\alpha=1/2$. There is a minority population of fluctuations with relatively larger $\lpar$ and $\lambda$ that do not appear to obey this rescaling, which suggests an imperfection of our model (unless it is a box-size convergence issue). However, this minority is small, which explains why it does not affect inertial-range scalings reported in Section \ref{sec:numsf}.  

In view of \eqref{pyglam}, the probability density function plotted in \figref{fig:lpardist} gives us an idea as to the shape of the function $f(y)$. Remarkably, at larger $y$, it scales very precisely as\footnote{\cite{zhdankin16intcy} appear to have observed a not entirely dissimilar scaling for the lengths and widths of Elsasser vorticity (current) sheets in their numerical simulations of MHD turbulence. These should correspond to our $\lpar$ and $\xi$ variables (which indeed have the same distribution for the most intense structures; see Section \ref{sec:sigma})---although \cite{zhdankin16intcy} have a completely different scheme for measuring them.}    
\beq
f(y)\sim \frac{1}{y^3}.
\label{eq:fytail}
\eeq
This is the ``fattest'' tail allowed by the condition \exref{eq:ordercond}, which we 
needed to be satisfied in order for our derivation of the $\hlam$ scaling of $P(q=0 | \hlam)$ 
in Section \ref{sec:alpha} to be valid. If the tail were any fatter, the integral 
in the numerator of \eqref{eq:Pq0lam} would be dominated by the upper limit and so the 
scaling of $P(q=0 | \hlam)$ with $\hlam$ would depend not just on $\alpha$, but also 
on the asymptotic form of $f(y)$. In view of the result \exref{eq:fytail}, our derivation 
survives, subject at most to a logarithmic correction (which, at this level of modelling, 
we view as irrelevant). It is an interesting question whether there is some compelling 
mechanism whereby the distribution of the anisotropy is allowed to be as broad as this 
but no broader.

\section{Discussion}
The model of strong Alfv\'enic turbulence presented in this paper leads to anisotropic scalings of the conditional structure functions in the local physical directions parallel to the local magnetic field, along the direction of the local fluctuation, and perpendicular to both of those, consistent with numerical evidence previously reported by \cite{mallet3d}. To achieve this, we have proposed four physically motivated conjectures: that the fluctuation amplitudes have an "anisotropic log-Poisson" distribution \citep[cf.][]{Chandran14,zhdankin16}, that the structures are sheetlike \citep[cf.][]{zhdankin13,zhdankin16intcy,Chandran14,howes2015}, that the critical-balance parameter (including dynamic alignment) is independent of scale \citep{rcb}, and finally that there is a constant flux of energy through parallel scales in the inertial range \citep[cf.][]{beresnyakanis}. This allows us to fix all the parameters of the model, resulting in simple predictions for the scalings of $n$th-order conditional structure functions in the perpendicular and parallel directions. In the fluctuation direction, we find the scalings approximately using an additional assumption. In all three directions, the scalings agree well with those previously reported in \cite{mallet3d} on the basis of numerical simulations (see Section \ref{sec:num}). Moreover, we find reasonable agreement between the distribution of parallel field increments in the numerically simulated turbulence and our log-Poisson model.

It is interesting to note that the predicted structure-function scalings in the perpendicular direction are nearly identical to those proposed by \cite{Chandran14}: the only difference being that in their model, the parameter $\beta\approx0.691$, whereas in our model in this paper, $\beta=1/\sqrt{2}\approx 0.707$. This is perhaps not too surprising: both models rely on a log-Poisson model for the fluctuation amplitude, refined critical balance, dynamic alignment, and constant flux of energy through scale. The differences are in certain details: firstly, in order to fix the scalings, \cite{Chandran14} constructed a detailed dynamical model for the collision of high-amplitude and low-amplitude fluctuations, whereas here we fixed the free parameters of our model via assumptions about the spatial dimensionality of the most intense fluctuations; secondly, for \cite{Chandran14}, the central quantity was the Elsasser-field increment at a given perpendicular point separation $\lambda$, whereas here most of the physical assumptions were about field increments at given parallel point separations $\lpar$\footnote{Both the numerical evidence (the log-Poisson fits in Section \ref{sec:numpdf}) and physical arguments based on the constant-flux and critical-balance assumptions (Section \ref{sec:poisson}) suggest that perhaps ``starting'' with parallel field increments in constructing theories of Alfv\'enic turbulence is a strategy that has some meaning.}---these assumptions were then used to model the ``RMHD turbulent ensemble'' of structures of random amplitudes and sizes in all three spatial directions. 

Having an explicit statistical model for such an ensemble then naturally allowed us to predict exponents not just in the perpendicular but also in the parallel and fluctuation directions. We have also been able to predict the distribution of the anisotropy between the parallel and perpendicular scales of the fluctuations: we have argued that, typically in the ensemble, $\lpar \sim \lambda^{1/2}$, independently of the fluctuation amplitude. This was indeed approximately the case in the numerical simulations reported by \cite{rcb} (see Section \ref{sec:anisdist}). 

The scalings that we have derived for the 2nd-order structure functions
[Eqs. (\ref{eq:zetaperp2}), (\ref{eq:zetapar2}) and (\ref{eq:zetafluc2})] and for the relationship between the perpendicular and parallel
coherence scales [Eqs. (\ref{eq:paraniseq}) and (\ref{eq:alpha})] are the same as proposed by
\citet{boldyrev} and, like his, are broadly based on the idea of alignment
between fluctuating fields. Note, however, that we have been able to
obtain these scalings without the need for the conjecture that the alignment
angle is equal to the "uncertainty" in the direction of the fluctuating field
($\theta \sim \delta z_\perp/v_{\rm A}$)---a conjecture that, if taken literally,
contradicts the scale invariance property of RMHD equations \citep{Beresnyak14}.
Here, this conjecture has been effectively replaced by Conjectures 1-3
(see sections \ref{sec:ensemble} and \ref{sec:fluxsheets}), leading to Eq. (\ref{eq:alpha}). Another important nuance is that
we are effectively assuming alignment between the Elsasser fields, rather
than between the velocity and magnetic fields
\citep[cf.][]{bl06,beresnyak09,mcatbolalign,Chandran14}---whereas
the two types of alignment are compatible and have been argued
to occur simultaneously \citep{pbxhel}, they are not
mutually necessary and fluctuations with one but not another have
been found in the solar wind \citep{wicksalign,wicks13}. Our theory cannot and
indeed does not predict velocity and magnetic-field scalings, which turn out to be
different both from each other and from the Elsasser fields (see section \ref{sec:numsf}).
 A more refined theory aspiring to explain this behaviour will have to address
the statistical dependence of the two Elsasser fields on each other
(i.e., the statistics of "local imbalance" in MHD turbulence).

It is a valid question whether scaling exponents computed from numerical simulations 
(e.g., the structure-function exponents in Section \ref{sec:numsf}) are believable if they are obtained without a systematic study of convergence in the limit of large Reynolds number [which would be 
the "gold standard" in hydrodynamics; see, e.g., \citet{gotoh2002}]. Indeed, in MHD 
turbulence such an approach was argued to be a sine qua non by Beresnyak \citep{Beresnyak14} and the question what to make of such studies at currently 
affordable --- possibly insufficient --- resolutions is a controversial one \citep{perez14,Beresnyak14}. We are not in position to carry out a numerical study   
that would exceed in size those already in existence and indeed we do not  
claim that aligned, 3D-anisotropic RMHD turbulent state that is seen in the numerical 
simulations at currently available resolutions will necessarily survive to arbitrarily 
small scales in the limit of $\rm{Re} \to \infty$. However, it appears that it does persist  
down to scale separations about a decade below the driving scale, 
which is the range captured in current simulations and, indeed, is not dramatically less broad 
than the universal part of the inertial range appears to be in the solar-wind turbulence. 
We therefore consider having a good model of Alfv\'enic fluctuations at these scales worthwhile 
and leave to future work the fascinating but still somewhat murky (and as yet difficult 
to address numerically) problem of what happens at even smaller scales.

It is fair to observe that in the solar wind, unlike in our model or in numerical simulations, definitive proof of a {\em scale-dependent} anisotropy in the inertial range between the perpendicular and fluctuation directions (i.e., scale-dependent alignment) has been elusive \citep{podestaalign,chen3d,wicksalign}, in contrast to the anisotropy with respect to the local mean field, which is quite well established \citep{horanis,podestaaniso,wicks10,chenmallet}. This could be due to the solar-wind expansion affecting the anisotropy \citep{verdini2015}, reflection of Alfv\'enic fluctuations close to the Sun \citep{perezchandran2013,heinemannolbert}, or the highly-imbalanced nature of the solar-wind turbulence \citep[e.g.,][]{wicks11,wicks13}. In any event, making quantitative contact between theory and data requires understanding of these effects. We lay no claim to such a complete understanding. Nevertheless, it appears that, by incorporating all three of intermittency, dynamic alignment, and critical balance in the same theoretical scheme, our model does at least help to make sense of the 3D-anisotropic statistics found in numerical simulations of homogeneous, balanced Alfv\'enic turbulence. 

\section*{Acknowledgements}
We are indebted to B.~D.~G.~Chandran for many important discussions, which substantially influenced this work. We also thank A. Beresnyak for useful comments, which helped improve our exposition. The work of A.M. was supported the NSF  under Award No. 1624501. The work of A.A.S. was supported in part by grants from UK  STFC and EPSRC. Simulations reported here used XSEDE, which is supported by the US NSF Grant ACI-1053575. Both authors thank the Wolfgang Pauli Institute, Vienna, where this work was conceived, for its hospitality. We also thank the anonymous referee for their helpful suggestions.
\bibliographystyle{mnras}
\bibliography{mainbib2} 

\begin{thebibliography}{}
\makeatletter
\relax
\def\mn@urlcharsother{\let\do\@makeother \do\$\do\&\do\#\do\^\do\_\do\%\do\~}
\def\mn@doi{\begingroup\mn@urlcharsother \@ifnextchar [ {\mn@doi@}
  {\mn@doi@[]}}
\def\mn@doi@[#1]#2{\def\@tempa{#1}\ifx\@tempa\@empty \href
  {http://dx.doi.org/#2} {doi:#2}\else \href {http://dx.doi.org/#2} {#1}\fi
  \endgroup}
\def\mn@eprint#1#2{\mn@eprint@#1:#2::\@nil}
\def\mn@eprint@arXiv#1{\href {http://arxiv.org/abs/#1} {{\tt arXiv:#1}}}
\def\mn@eprint@dblp#1{\href {http://dblp.uni-trier.de/rec/bibtex/#1.xml}
  {dblp:#1}}
\def\mn@eprint@#1:#2:#3:#4\@nil{\def\@tempa {#1}\def\@tempb {#2}\def\@tempc
  {#3}\ifx \@tempc \@empty \let \@tempc \@tempb \let \@tempb \@tempa \fi \ifx
  \@tempb \@empty \def\@tempb {arXiv}\fi \@ifundefined
  {mn@eprint@\@tempb}{\@tempb:\@tempc}{\expandafter \expandafter \csname
  mn@eprint@\@tempb\endcsname \expandafter{\@tempc}}}

\bibitem[\protect\citeauthoryear{{Beresnyak}}{{Beresnyak}}{2014}]{Beresnyak14}
{Beresnyak} A.,  2014, \mn@doi [\apjl] {10.1088/2041-8205/784/2/L20}, \href
  {http://adsabs.harvard.edu/abs/2014ApJ...784L..20B} {784, L20}

\bibitem[\protect\citeauthoryear{{Beresnyak}}{{Beresnyak}}{2015}]{beresnyakanis}
{Beresnyak} A.,  2015, \mn@doi [\apjl] {10.1088/2041-8205/801/1/L9}, \href
  {http://adsabs.harvard.edu/abs/2015ApJ...801L...9B} {801, L9}

\bibitem[\protect\citeauthoryear{{Beresnyak} \& {Lazarian}}{{Beresnyak} \&
  {Lazarian}}{2006}]{bl06}
{Beresnyak} A.,  {Lazarian} A.,  2006, \mn@doi [\apjl] {10.1086/503708}, \href
  {http://adsabs.harvard.edu/abs/2006ApJ...640L.175B} {640, L175}

\bibitem[\protect\citeauthoryear{{Beresnyak} \& {Lazarian}}{{Beresnyak} \&
  {Lazarian}}{2009}]{beresnyak09}
{Beresnyak} A.,  {Lazarian} A.,  2009, \mn@doi [\apj]
  {10.1088/0004-637X/702/2/1190}, \href
  {http://adsabs.harvard.edu/abs/2009ApJ...702.1190B} {702, 1190}

\bibitem[\protect\citeauthoryear{{Bigot}, {Galtier}  \& {Politano}}{{Bigot}
  et~al.}{2008}]{bigot08}
{Bigot} B.,  {Galtier} S.,   {Politano} H.,  2008, \mn@doi [\pre]
  {10.1103/PhysRevE.78.066301}, \href
  {http://adsabs.harvard.edu/abs/2008PhRvE..78f6301B} {78, 066301}

\bibitem[\protect\citeauthoryear{{Boldyrev}}{{Boldyrev}}{2006}]{boldyrev}
{Boldyrev} S.,  2006, \mn@doi [\prl] {10.1103/PhysRevLett.96.115002}, \href
  {http://adsabs.harvard.edu/abs/2006PhRvL..96k5002B} {96, 115002}

\bibitem[\protect\citeauthoryear{{Bruno} \& {Carbone}}{{Bruno} \&
  {Carbone}}{2013}]{bruno2013}
{Bruno} R.,  {Carbone} V.,  2013, \mn@doi [Living Rev. Solar Phys.]
  {10.12942/lrsp-2013-2}, \href
  {http://adsabs.harvard.edu/abs/2013LRSP...10....2B} {10, 2}

\bibitem[\protect\citeauthoryear{{Chandran}, {Schekochihin}  \&
  {Mallet}}{{Chandran} et~al.}{2015}]{Chandran14}
{Chandran} B.~D.~G.,  {Schekochihin} A.~A.,   {Mallet} A.,  2015, \mn@doi
  [\apj] {10.1088/0004-637X/807/1/39}, \href
  {http://adsabs.harvard.edu/abs/2015ApJ...807...39C} {807, 39}

\bibitem[\protect\citeauthoryear{{Chasapis} et~al.,}{{Chasapis}
  et~al.}{2015}]{chasapis15}
{Chasapis} A.,  et~al., 2015, \mn@doi [\apjl] {10.1088/2041-8205/804/1/L1},
  \href {http://adsabs.harvard.edu/abs/2015ApJ...804L...1C} {804, L1}

\bibitem[\protect\citeauthoryear{{Chen}, {Mallet}, {Yousef}, {Schekochihin}  \&
  {Horbury}}{{Chen} et~al.}{2011}]{chenmallet}
{Chen} C.~H.~K.,  {Mallet} A.,  {Yousef} T.~A.,  {Schekochihin} A.~A.,
  {Horbury} T.~S.,  2011, \mn@doi [\mnras] {10.1111/j.1365-2966.2011.18933.x},
  \href {http://adsabs.harvard.edu/abs/2011MNRAS.415.3219C} {415, 3219}

\bibitem[\protect\citeauthoryear{{Chen}, {Mallet}, {Schekochihin}, {Horbury},
  {Wicks}  \& {Bale}}{{Chen} et~al.}{2012}]{chen3d}
{Chen} C.~H.~K.,  {Mallet} A.,  {Schekochihin} A.~A.,  {Horbury} T.~S.,
  {Wicks} R.~T.,   {Bale} S.~D.,  2012, \mn@doi [\apj]
  {10.1088/0004-637X/758/2/120}, \href
  {http://adsabs.harvard.edu/abs/2012ApJ...758..120C} {758, 120}

\bibitem[\protect\citeauthoryear{{Cho} \& {Vishniac}}{{Cho} \&
  {Vishniac}}{2000}]{chovishniac}
{Cho} J.,  {Vishniac} E.~T.,  2000, \mn@doi [\apj] {10.1086/309213}, \href
  {http://adsabs.harvard.edu/abs/2000ApJ...539..273C} {539, 273}

\bibitem[\protect\citeauthoryear{{Dubrulle}}{{Dubrulle}}{1994}]{dubrulle94}
{Dubrulle} B.,  1994, \mn@doi [\prl] {10.1103/PhysRevLett.73.959}, \href
  {http://adsabs.harvard.edu/abs/1994PhRvL..73..959D} {73, 959}

\bibitem[\protect\citeauthoryear{{Frisch}}{{Frisch}}{1995}]{frischbook}
{Frisch} U.,  1995, {Turbulence: The Legacy of A. N. Kolmogorov}.
Cambridge University Press

\bibitem[\protect\citeauthoryear{{Goldreich} \& {Sridhar}}{{Goldreich} \&
  {Sridhar}}{1995}]{gs95}
{Goldreich} P.,  {Sridhar} S.,  1995, \mn@doi [\apj] {10.1086/175121}, \href
  {http://adsabs.harvard.edu/abs/1995ApJ...438..763G} {438, 763}

\bibitem[\protect\citeauthoryear{{Goldreich} \& {Sridhar}}{{Goldreich} \&
  {Sridhar}}{1997}]{gs97}
{Goldreich} P.,  {Sridhar} S.,  1997, \apj, \href
  {http://adsabs.harvard.edu/abs/1997ApJ...485..680G} {485, 680}

\bibitem[\protect\citeauthoryear{{Gotoh}, {Fukayama}  \& {Nakano}}{{Gotoh}
  et~al.}{2002}]{gotoh2002}
{Gotoh} T.,  {Fukayama} D.,   {Nakano} T.,  2002, \mn@doi [Physics of Fluids]
  {10.1063/1.1448296}, \href
  {http://adsabs.harvard.edu/abs/2002PhFl...14.1065G} {14, 1065}

\bibitem[\protect\citeauthoryear{{Greco}, {Matthaeus}, {Servidio}, {Chuychai}
  \& {Dmitruk}}{{Greco} et~al.}{2009}]{greco09}
{Greco} A.,  {Matthaeus} W.~H.,  {Servidio} S.,  {Chuychai} P.,   {Dmitruk} P.,
   2009, \mn@doi [\apjl] {10.1088/0004-637X/691/2/L111}, \href
  {http://adsabs.harvard.edu/abs/2009ApJ...691L.111G} {691, L111}

\bibitem[\protect\citeauthoryear{{Heinemann} \& {Olbert}}{{Heinemann} \&
  {Olbert}}{1980}]{heinemannolbert}
{Heinemann} M.,  {Olbert} S.,  1980, \mn@doi [\jgr] {10.1029/JA085iA03p01311},
  \href {http://adsabs.harvard.edu/abs/1980JGR....85.1311H} {85, 1311}

\bibitem[\protect\citeauthoryear{{Horbury}, {Forman}  \& {Oughton}}{{Horbury}
  et~al.}{2008}]{horanis}
{Horbury} T.~S.,  {Forman} M.,   {Oughton} S.,  2008, \mn@doi [\prl]
  {10.1103/PhysRevLett.101.175005}, \href
  {http://adsabs.harvard.edu/abs/2008PhRvL.101q5005H} {101, 175005}

\bibitem[\protect\citeauthoryear{{Howes}}{{Howes}}{2015}]{howes2015}
{Howes} G.~G.,  2015, \mn@doi [Phil. Trans. R. Soc. Lond. A]
  {10.1098/rsta.2014.0145}, \href
  {http://adsabs.harvard.edu/abs/2015RSPTA.37340145H} {373, 20140145}

\bibitem[\protect\citeauthoryear{{Kolmogorov}}{{Kolmogorov}}{1941}]{k41}
{Kolmogorov} A.,  1941, Dokl. Akad. Nauk. SSSR, 30, 301

\bibitem[\protect\citeauthoryear{{Kolmogorov}}{{Kolmogorov}}{1962}]{k62}
{Kolmogorov} A.~N.,  1962, \mn@doi [J.\ Fluid Mech.]
  {10.1017/S0022112062000518}, \href
  {http://adsabs.harvard.edu/abs/1962JFM....13...82K} {13, 82}

\bibitem[\protect\citeauthoryear{{Mallet}, {Schekochihin}  \&
  {Chandran}}{{Mallet} et~al.}{2015}]{rcb}
{Mallet} A.,  {Schekochihin} A.~A.,   {Chandran} B.~D.~G.,  2015, \mn@doi
  [\mnras] {10.1093/mnrasl/slv021}, \href
  {http://adsabs.harvard.edu/abs/2015MNRAS.449L..77M} {449, L77}

\bibitem[\protect\citeauthoryear{{Mallet}, {Schekochihin}, {Chandran}, {Chen},
  {Horbury}, {Wicks}  \& {Greenan}}{{Mallet} et~al.}{2016}]{mallet3d}
{Mallet} A.,  {Schekochihin} A.~A.,  {Chandran} B.~D.~G.,  {Chen} C.~H.~K.,
  {Horbury} T.~S.,  {Wicks} R.~T.,   {Greenan} C.~C.,  2016, \mn@doi [\mnras]
  {10.1093/mnras/stw802}, \href
  {http://adsabs.harvard.edu/abs/2016MNRAS.459.2130M} {459, 2130}

\bibitem[\protect\citeauthoryear{{Maron} \& {Goldreich}}{{Maron} \&
  {Goldreich}}{2001}]{marongoldreich}
{Maron} J.,  {Goldreich} P.,  2001, \mn@doi [\apj] {10.1086/321413}, \href
  {http://adsabs.harvard.edu/abs/2001ApJ...554.1175M} {554, 1175}

\bibitem[\protect\citeauthoryear{{Mason}, {Cattaneo}  \& {Boldyrev}}{{Mason}
  et~al.}{2006}]{mcatbolalign}
{Mason} J.,  {Cattaneo} F.,   {Boldyrev} S.,  2006, \mn@doi [\prl]
  {10.1103/PhysRevLett.97.255002}, \href
  {http://adsabs.harvard.edu/abs/2006PhRvL..97y5002M} {97, 255002}

\bibitem[\protect\citeauthoryear{{Matthaeus}, {Ghosh}, {Oughton}  \&
  {Roberts}}{{Matthaeus} et~al.}{1996}]{matthaeus96}
{Matthaeus} W.~H.,  {Ghosh} S.,  {Oughton} S.,   {Roberts} D.~A.,  1996,
  \mn@doi [\jgr] {10.1029/95JA03830}, \href
  {http://adsabs.harvard.edu/abs/1996JGR...101.7619M} {101, 7619}

\bibitem[\protect\citeauthoryear{{Matthaeus}, {Oughton}, {Ghosh}  \&
  {Hossain}}{{Matthaeus} et~al.}{1998}]{matthaeus98}
{Matthaeus} W.~H.,  {Oughton} S.,  {Ghosh} S.,   {Hossain} M.,  1998, \mn@doi
  [\prl] {10.1103/PhysRevLett.81.2056}, \href
  {http://adsabs.harvard.edu/abs/1998PhRvL..81.2056M} {81, 2056}

\bibitem[\protect\citeauthoryear{{Matthaeus}, {Servidio}, {Dmitruk}, {Carbone},
  {Oughton}, {Wan}  \& {Osman}}{{Matthaeus} et~al.}{2012}]{Matthaeus12}
{Matthaeus} W.~H.,  {Servidio} S.,  {Dmitruk} P.,  {Carbone} V.,  {Oughton} S.,
   {Wan} M.,   {Osman} K.~T.,  2012, \mn@doi [\apj]
  {10.1088/0004-637X/750/2/103}, \href
  {http://adsabs.harvard.edu/abs/2012ApJ...750..103M} {750, 103}

\bibitem[\protect\citeauthoryear{{M{\"u}ller}, {Biskamp}  \&
  {Grappin}}{{M{\"u}ller} et~al.}{2003}]{mullerbiskamp}
{M{\"u}ller} W.-C.,  {Biskamp} D.,   {Grappin} R.,  2003, \mn@doi [\pre]
  {10.1103/PhysRevE.67.066302}, \href
  {http://adsabs.harvard.edu/abs/2003PhRvE..67f6302M} {67, 066302}

\bibitem[\protect\citeauthoryear{{Osman}, {Matthaeus}, {Gosling}, {Greco},
  {Servidio}, {Hnat}, {Chapman}  \& {Phan}}{{Osman} et~al.}{2014}]{osman14rec}
{Osman} K.~T.,  {Matthaeus} W.~H.,  {Gosling} J.~T.,  {Greco} A.,  {Servidio}
  S.,  {Hnat} B.,  {Chapman} S.~C.,   {Phan} T.~D.,  2014, \mn@doi [\prl]
  {10.1103/PhysRevLett.112.215002}, \href
  {http://adsabs.harvard.edu/abs/2014PhRvL.112u5002O} {112, 215002}

\bibitem[\protect\citeauthoryear{{Oughton}, {Priest}  \& {Matthaeus}}{{Oughton}
  et~al.}{1994}]{oughton94}
{Oughton} S.,  {Priest} E.~R.,   {Matthaeus} W.~H.,  1994, \mn@doi [J. Fluid
  Mech.] {10.1017/S0022112094002867}, \href
  {http://adsabs.harvard.edu/abs/1994JFM...280...95O} {280, 95}

\bibitem[\protect\citeauthoryear{{Oughton}, {Dmitruk}  \&
  {Matthaeus}}{{Oughton} et~al.}{2004}]{oughton04}
{Oughton} S.,  {Dmitruk} P.,   {Matthaeus} W.~H.,  2004, \mn@doi [Phys.
  Plasmas] {10.1063/1.1705652}, \href
  {http://adsabs.harvard.edu/abs/2004PhPl...11.2214O} {11, 2214}

\bibitem[\protect\citeauthoryear{{Perez} \& {Boldyrev}}{{Perez} \&
  {Boldyrev}}{2009}]{pbxhel}
{Perez} J.~C.,  {Boldyrev} S.,  2009, \mn@doi [\prl]
  {10.1103/PhysRevLett.102.025003}, \href
  {http://adsabs.harvard.edu/abs/2009PhRvL.102b5003P} {102, 025003}

\bibitem[\protect\citeauthoryear{{Perez} \& {Chandran}}{{Perez} \&
  {Chandran}}{2013}]{perezchandran2013}
{Perez} J.~C.,  {Chandran} B.~D.~G.,  2013, \mn@doi [\apj]
  {10.1088/0004-637X/776/2/124}, \href
  {http://adsabs.harvard.edu/abs/2013ApJ...776..124P} {776, 124}

\bibitem[\protect\citeauthoryear{{Perez}, {Mason}, {Boldyrev}  \&
  {Cattaneo}}{{Perez} et~al.}{2014}]{perez14}
{Perez} J.~C.,  {Mason} J.,  {Boldyrev} S.,   {Cattaneo} F.,  2014, \mn@doi
  [\apjl] {10.1088/2041-8205/793/1/L13}, \href
  {http://adsabs.harvard.edu/abs/2014ApJ...793L..13P} {793, L13}

\bibitem[\protect\citeauthoryear{{Perri}, {Goldstein}, {Dorelli}  \&
  {Sahraoui}}{{Perri} et~al.}{2012}]{perri12}
{Perri} S.,  {Goldstein} M.~L.,  {Dorelli} J.~C.,   {Sahraoui} F.,  2012,
  \mn@doi [\prl] {10.1103/PhysRevLett.109.191101}, \href
  {http://adsabs.harvard.edu/abs/2012PhRvL.109s1101P} {109, 191101}

\bibitem[\protect\citeauthoryear{{Podesta}}{{Podesta}}{2009}]{podestaaniso}
{Podesta} J.~J.,  2009, \mn@doi [\apj] {10.1088/0004-637X/698/2/986}, \href
  {http://adsabs.harvard.edu/abs/2009ApJ...698..986P} {698, 986}

\bibitem[\protect\citeauthoryear{{Podesta}, {Chandran}, {Bhattacharjee},
  {Roberts}  \& {Goldstein}}{{Podesta} et~al.}{2009}]{podestaalign}
{Podesta} J.~J.,  {Chandran} B.~D.~G.,  {Bhattacharjee} A.,  {Roberts} D.~A.,
  {Goldstein} M.~L.,  2009, \mn@doi [\jgr] {10.1029/2008JA013504}, \href
  {http://adsabs.harvard.edu/abs/2009JGRA..114.1107P} {114, 1107}

\bibitem[\protect\citeauthoryear{{Sato}}{{Sato}}{2013}]{satobook}
{Sato} K.,  2013, {L\'evy Processes and Infinitely Divisible Distributions}.
Cambdridge University Press

\bibitem[\protect\citeauthoryear{{Schekochihin}, {Cowley}, {Dorland},
  {Hammett}, {Howes}, {Quataert}  \& {Tatsuno}}{{Schekochihin}
  et~al.}{2009}]{schektome2009}
{Schekochihin} A.~A.,  {Cowley} S.~C.,  {Dorland} W.,  {Hammett} G.~W.,
  {Howes} G.~G.,  {Quataert} E.,   {Tatsuno} T.,  2009, \mn@doi [\apjs]
  {10.1088/0067-0049/182/1/310}, \href
  {http://adsabs.harvard.edu/abs/2009ApJS..182..310S} {182, 310}

\bibitem[\protect\citeauthoryear{{She} \& {Leveque}}{{She} \&
  {Leveque}}{1994}]{sl94}
{She} Z.-S.,  {Leveque} E.,  1994, \mn@doi [\prl] {10.1103/PhysRevLett.72.336},
  \href {http://adsabs.harvard.edu/abs/1994PhRvL..72..336S} {72, 336}

\bibitem[\protect\citeauthoryear{{She} \& {Waymire}}{{She} \&
  {Waymire}}{1995}]{shewaymire}
{She} Z.-S.,  {Waymire} E.~C.,  1995, \mn@doi [\prl]
  {10.1103/PhysRevLett.74.262}, \href
  {http://adsabs.harvard.edu/abs/1995PhRvL..74..262S} {74, 262}

\bibitem[\protect\citeauthoryear{{Shebalin}, {Matthaeus}  \&
  {Montgomery}}{{Shebalin} et~al.}{1983}]{shebalin83}
{Shebalin} J.~V.,  {Matthaeus} W.~H.,   {Montgomery} D.,  1983, \mn@doi [J.
  Plasma Phys.] {10.1017/S0022377800000933}, \href
  {http://adsabs.harvard.edu/abs/1983JPlPh..29..525S} {29, 525}

\bibitem[\protect\citeauthoryear{{Verdini} \& {Grappin}}{{Verdini} \&
  {Grappin}}{2015}]{verdini2015}
{Verdini} A.,  {Grappin} R.,  2015, \mn@doi [\apjl]
  {10.1088/2041-8205/808/2/L34}, \href
  {http://adsabs.harvard.edu/abs/2015ApJ...808L..34V} {808, L34}

\bibitem[\protect\citeauthoryear{{Wicks}, {Horbury}, {Chen}  \&
  {Schekochihin}}{{Wicks} et~al.}{2010}]{wicks10}
{Wicks} R.~T.,  {Horbury} T.~S.,  {Chen} C.~H.~K.,   {Schekochihin} A.~A.,
  2010, \mn@doi [\mnras] {10.1111/j.1745-3933.2010.00898.x}, \href
  {http://adsabs.harvard.edu/abs/2010MNRAS.407L..31W} {407, L31}

\bibitem[\protect\citeauthoryear{{Wicks}, {Horbury}, {Chen}  \&
  {Schekochihin}}{{Wicks} et~al.}{2011}]{wicks11}
{Wicks} R.~T.,  {Horbury} T.~S.,  {Chen} C.~H.~K.,   {Schekochihin} A.~A.,
  2011, \mn@doi [\prl] {10.1103/PhysRevLett.106.045001}, \href
  {http://adsabs.harvard.edu/abs/2011PhRvL.106d5001W} {106, 045001}

\bibitem[\protect\citeauthoryear{{Wicks}, {Mallet}, {Horbury}, {Chen},
  {Schekochihin}  \& {Mitchell}}{{Wicks} et~al.}{2013a}]{wicksalign}
{Wicks} R.~T.,  {Mallet} A.,  {Horbury} T.~S.,  {Chen} C.~H.~K.,
  {Schekochihin} A.~A.,   {Mitchell} J.~J.,  2013a, \mn@doi [\prl]
  {10.1103/PhysRevLett.110.025003}, \href
  {http://adsabs.harvard.edu/abs/2013PhRvL.110b5003W} {110, 025003}

\bibitem[\protect\citeauthoryear{{Wicks}, {Roberts}, {Mallet}, {Schekochihin},
  {Horbury}  \& {Chen}}{{Wicks} et~al.}{2013b}]{wicks13}
{Wicks} R.~T.,  {Roberts} D.~A.,  {Mallet} A.,  {Schekochihin} A.~A.,
  {Horbury} T.~S.,   {Chen} C.~H.~K.,  2013b, \mn@doi [\apj]
  {10.1088/0004-637X/778/2/177}, \href
  {http://adsabs.harvard.edu/abs/2013ApJ...778..177W} {778, 177}

\bibitem[\protect\citeauthoryear{{Zhdankin}, {Uzdensky}, {Perez}  \&
  {Boldyrev}}{{Zhdankin} et~al.}{2013}]{zhdankin13}
{Zhdankin} V.,  {Uzdensky} D.~A.,  {Perez} J.~C.,   {Boldyrev} S.,  2013,
  \mn@doi [\apj] {10.1088/0004-637X/771/2/124}, \href
  {http://adsabs.harvard.edu/abs/2013ApJ...771..124Z} {771, 124}

\bibitem[\protect\citeauthoryear{{Zhdankin}, {Boldyrev}  \&
  {Uzdensky}}{{Zhdankin} et~al.}{2016a}]{zhdankin16intcy}
{Zhdankin} V.,  {Boldyrev} S.,   {Uzdensky} D.~A.,  2016a, \mn@doi [Phys.\
  Plasmas] {10.1063/1.4944820}, \href
  {http://adsabs.harvard.edu/abs/2016PhPl...23e5705Z} {23, 055705}

\bibitem[\protect\citeauthoryear{{Zhdankin}, {Boldyrev}  \& {Chen}}{{Zhdankin}
  et~al.}{2016b}]{zhdankin16}
{Zhdankin} V.,  {Boldyrev} S.,   {Chen} C.~H.~K.,  2016b, \mn@doi [\mnras]
  {10.1093/mnrasl/slv208}, \href
  {http://adsabs.harvard.edu/abs/2016MNRAS.457L..69Z} {457, L69}

\makeatother
\end{thebibliography}


\bsp	
\label{lastpage}
\end{document}